\documentclass[twocolumn,aps,prc,floatfix,superscriptaddress]{revtex4-1}

\usepackage{graphicx}
\usepackage{dcolumn}
\usepackage{amsmath}
\usepackage{supertabular}
\usepackage{multirow}
\usepackage{multicols}
\RequirePackage{lineno}
\usepackage{color}
\usepackage{cancel}
\usepackage{ulem}
\usepackage{array}
\usepackage{tabularx}

\newcommand{\snn}{\ensuremath{\sqrt{s_{_{\rm NN}}}}}

\newcommand{\caab}{\ensuremath{C_{1,1,2}}}
\newcommand{\cabc}{\ensuremath{C_{1,2,3}}}
\newcommand{\cacd}{\ensuremath{C_{1,3,4}}}
\newcommand{\cbbd}{\ensuremath{C_{2,2,4}}}
\newcommand{\cbce}{\ensuremath{C_{2,3,5}}}
\newcommand{\cccf}{\ensuremath{C_{3,3,6}}}
\newcommand{\ccdg}{\ensuremath{C_{3,4,7}}}
\newcommand{\cijk}{\ensuremath{C_{m,n,m+n}}}

\newcommand{\npart}{\ensuremath{N_{\mathrm{part}}}}
\newcommand{\npartsq}{\ensuremath{N_{\mathrm{part}}^{2}}}

\begin{document}

\title{ Harmonic decomposition of three-particle azimuthal correlations at RHIC }

\affiliation{AGH University of Science and Technology, FPACS, Cracow 30-059, Poland}
\affiliation{Argonne National Laboratory, Argonne, Illinois 60439}
\affiliation{Brookhaven National Laboratory, Upton, New York 11973}
\affiliation{University of California, Berkeley, California 94720}
\affiliation{University of California, Davis, California 95616}
\affiliation{University of California, Los Angeles, California 90095}
\affiliation{Central China Normal University, Wuhan, Hubei 430079}
\affiliation{University of Illinois at Chicago, Chicago, Illinois 60607}
\affiliation{Creighton University, Omaha, Nebraska 68178}
\affiliation{Czech Technical University in Prague, FNSPE, Prague, 115 19, Czech Republic}
\affiliation{Nuclear Physics Institute AS CR, 250 68 Prague, Czech Republic}
\affiliation{Frankfurt Institute for Advanced Studies FIAS, Frankfurt 60438, Germany}
\affiliation{Institute of Physics, Bhubaneswar 751005, India}
\affiliation{Indiana University, Bloomington, Indiana 47408}
\affiliation{Alikhanov Institute for Theoretical and Experimental Physics, Moscow 117218, Russia}
\affiliation{University of Jammu, Jammu 180001, India}
\affiliation{Joint Institute for Nuclear Research, Dubna, 141 980, Russia}
\affiliation{Kent State University, Kent, Ohio 44242}
\affiliation{University of Kentucky, Lexington, Kentucky, 40506-0055}
\affiliation{Lamar University, Physics Department, Beaumont, Texas 77710}
\affiliation{Institute of Modern Physics, Chinese Academy of Sciences, Lanzhou, Gansu 730000}
\affiliation{Lawrence Berkeley National Laboratory, Berkeley, California 94720}
\affiliation{Lehigh University, Bethlehem, PA, 18015}
\affiliation{Max-Planck-Institut fur Physik, Munich 80805, Germany}
\affiliation{Michigan State University, East Lansing, Michigan 48824}
\affiliation{National Research Nuclear University MEPhI, Moscow 115409, Russia}
\affiliation{National Institute of Science Education and Research, Bhubaneswar 751005, India}
\affiliation{National Cheng Kung University, Tainan 70101 }
\affiliation{Ohio State University, Columbus, Ohio 43210}
\affiliation{Institute of Nuclear Physics PAN, Cracow 31-342, Poland}
\affiliation{Panjab University, Chandigarh 160014, India}
\affiliation{Pennsylvania State University, University Park, Pennsylvania 16802}
\affiliation{Institute of High Energy Physics, Protvino 142281, Russia}
\affiliation{Purdue University, West Lafayette, Indiana 47907}
\affiliation{Pusan National University, Pusan 46241, Korea}
\affiliation{Rice University, Houston, Texas 77251}
\affiliation{University of Science and Technology of China, Hefei, Anhui 230026}
\affiliation{Shandong University, Jinan, Shandong 250100}
\affiliation{Shanghai Institute of Applied Physics, Chinese Academy of Sciences, Shanghai 201800}
\affiliation{State University Of New York, Stony Brook, NY 11794}
\affiliation{Temple University, Philadelphia, Pennsylvania 19122}
\affiliation{Texas A\&M University, College Station, Texas 77843}
\affiliation{University of Texas, Austin, Texas 78712}
\affiliation{University of Houston, Houston, Texas 77204}
\affiliation{Tsinghua University, Beijing 100084}
\affiliation{University of Tsukuba, Tsukuba, Ibaraki, Japan,}
\affiliation{Southern Connecticut State University, New Haven, CT, 06515}
\affiliation{United States Naval Academy, Annapolis, Maryland, 21402}
\affiliation{Valparaiso University, Valparaiso, Indiana 46383}
\affiliation{Variable Energy Cyclotron Centre, Kolkata 700064, India}
\affiliation{Warsaw University of Technology, Warsaw 00-661, Poland}
\affiliation{Wayne State University, Detroit, Michigan 48201}
\affiliation{World Laboratory for Cosmology and Particle Physics (WLCAPP), Cairo 11571, Egypt}
\affiliation{Yale University, New Haven, Connecticut 06520}

\author{L.~Adamczyk}\affiliation{AGH University of Science and Technology, FPACS, Cracow 30-059, Poland}
\author{J.~K.~Adkins}\affiliation{University of Kentucky, Lexington, Kentucky, 40506-0055}
\author{G.~Agakishiev}\affiliation{Joint Institute for Nuclear Research, Dubna, 141 980, Russia}
\author{M.~M.~Aggarwal}\affiliation{Panjab University, Chandigarh 160014, India}
\author{Z.~Ahammed}\affiliation{Variable Energy Cyclotron Centre, Kolkata 700064, India}
\author{N.~N.~Ajitanand}\affiliation{State University Of New York, Stony Brook, NY 11794}
\author{I.~Alekseev}\affiliation{Alikhanov Institute for Theoretical and Experimental Physics, Moscow 117218, Russia}\affiliation{National Research Nuclear University MEPhI, Moscow 115409, Russia}
\author{D.~M.~Anderson}\affiliation{Texas A\&M University, College Station, Texas 77843}
\author{R.~Aoyama}\affiliation{University of Tsukuba, Tsukuba, Ibaraki, Japan,}
\author{A.~Aparin}\affiliation{Joint Institute for Nuclear Research, Dubna, 141 980, Russia}
% ...

\collaboration{STAR Collaboration}\noaffiliation

\date{\today}

\begin{abstract} %
We present measurements of three-particle correlations for
various harmonics in Au+Au collisions at energies ranging from
$\sqrt{s_{{\rm NN}}}=7.7$ to 200 GeV using the STAR detector. The quantity
$\langle\cos(m\phi_1+n\phi_2-(m+n)\phi_3)\rangle$ is evaluated as a
function of $\sqrt{s_{{\rm NN}}}$, collision centrality, transverse momentum,
$p_T$, pseudo-rapidity difference, $\Delta\eta$, and harmonics ($m$ and
$n$). These data provide detailed information on global event properties like the three dimensional structure of the initial overlap
region, the expansion dynamics of the matter produced in the
collisions, and the transport properties of the medium. A strong dependence on $\Delta\eta$ is observed for most harmonic combinations consistent with breaking of longitudinal boost invariance.
Data reveal changes with energy in the two-particle correlation
functions relative to the second-harmonic event-plane and
provide ways to constrain models of heavy-ion collisions over a wide
range of collision energies.
 
\end{abstract}


\maketitle


\section{Introduction}

Heavy nuclei are collided at facilities like the Relativistic Heavy
Ion Collider (RHIC) and the Large Hadron Collider (LHC) in order to
study the emergent properties of matter with quarks and gluons as the
dominant degrees-of-freedom: a quark-gluon plasma
(QGP)~\cite{Collins:1974ky,Chin:1978gj,Kapusta:1979fh,Anishetty:1980zp}. The QGP
is a form of matter that existed in the early universe when its
ambient temperature was more than 155 MeV or 200 thousand times hotter
than the center of the
sun~\cite{Borsanyi:2010bp,Bhattacharya:2014ara}. As temperatures drop,
quarks and gluons no longer possess the energy necessary to overcome
the confining forces of QCD and they become confined into color
neutral hadrons and the QGP transitions smoothly and continuously into
a gas of hadrons~\cite{Aoki:2006we}. This transition occurred in the
early universe at about one microsecond after the big bang. Heavy-ion
collisions provide the only known method to recreate and study that
phase transition in a laboratory setting.

To provide the clearest
possible picture of this phase transition, a beam energy scan was
carried out at RHIC with collision energies ranging from
$\sqrt{s_{{\rm NN}}}$=200 GeV down to 7.7 GeV. Lowering the beam energy
naturally reduces the initial temperature of the matter created in the
collisions providing information on how the transport properties and equilibrium of
the matter vary with temperature~\cite{Aggarwal:2010cw}.
These heavy-ion collisions however create systems that are both very
small and short-lived. The characteristic size of the collision region
is the size of a nucleus or approximately $10^{-14}$ meters
across. This system expands in the longitudinal direction and
eventually in the transverse direction so that the energy density
drops quickly. Any quark gluon plasma that exists will only survive
for on the order $5\times10^{-23}$ seconds. Given the smallness of the
system and its very brief lifetime, it is challenging to determine the
nature of the matter left behind after the initial
collisions. Physicists rely on indirect observations based on
particles streaming from the collision region which are observed long
after any QGP has ceased to exist.  Correlations between
these produced particles have provided insight into the early phases
of the expansion as well as the characteristics of the matter
undergoing the expansion~\cite{reviews}. The dependence of the
correlations on the azimuthal angle between particles
$\Delta\phi=\phi_1-\phi_2$ has proven to be particularly
informative. Data have revealed that even when particle pairs are
separated by large angles in the longitudinal direction (large
$\Delta\eta$), they remain strongly correlated in the azimuthal
direction. This correlation manifests itself as a prominent ridge-like
structure in two-particle, $\Delta\eta$, $\Delta\phi$, correlation
functions~\cite{ridgedata}. The origin of this ridge has been traced
to the initial geometry of the collision region where flux tubes are
localized in the transverse direction but stretch over a long distance
in the longitudinal
direction~\cite{radflow,Mishra:2007tw,Sorensen:2008dm,Takahashi:2009na}. How
well these structures from the initial geometry are translated into
correlations between particles emitted from the collision region
reveals information about the medium's viscosity: the larger
the viscosity, the more washed out the correlations will
become~\cite{Sorensen:2011hm}. To study these effects, it is convenient
to examine the coefficients of a Fourier transform of the $\Delta\phi$
dependence of the two-particle correlation
functions~\cite{earlyv3}. These coefficients have been variously
labeled as $V_{n}$, $a_{n}$, or $v_{n}^{2}\{2\}$ where $n$ is the harmonic. Although the latter
is perhaps more cumbersome, we have maintained its usage owing to
its connection to the original terminology used for two-particle cumulants which has been in use for more than a decade~\cite{Adler:2002pu}. While $v_{n}^{2}\{2\}=\langle\cos n(\Delta\phi)\rangle$ has been studied as a function of $\sqrt{s_{{\rm NN}}}$, centrality, harmonic $n$, $p_T$, and $\Delta\eta$~\cite{besv3}, in this paper we extend this analysis from two-particle correlations to three-particle mixed harmonic correlations of the form $\langle \cos(m\phi_1+n\phi_2-(m+n)\phi_3)\rangle$~\cite{Bhalerao:2013ina} where $m$ and $n$ are positive integers.

Extending the analysis of azimuthal correlations from two to three
particles provides several benefits. First, the three particle
correlations provide greater sensitivity to the three-dimensional
structure of the initial state by for example revealing information about the two-particle $\Delta\eta-\Delta\phi$ correlations with respect to the reaction plane. Many models of heavy-ion collisions
make the simplifying assumption that the initial geometry of the
collision overlap does not vary with rapidity and that a boost
invariant central rapidity plateau may be
considered~\cite{Bjorken:1982qr}. It is likely however that this
assumption is broken by the asymmetric nature of the initial state and
that precision comparisons between models and data will require a
better understanding of the initial state fluctuations in all three
dimensions~\cite{Denicol:2015nhu}. Second, the new measurements can
constrain
models~\cite{Teaney:2010vd,Qiu:2012uy,Niemi:2015qia,Teaney:2013dta}. When
signals seen in two-particle correlations may be mocked up by multiple
effects, three-particle correlations can break those ambiguities. This
is important as models become more sophisticated by including for
example bulk viscosity, shear viscosity, and their temperature
dependence~\cite{Ryu:2015vwa}. Also, three-particle correlations can
reveal information about how two-particle correlations change as a
function of their angle with respect to the reaction plane. When one
of the harmonics $m$, $n$, or $m+n$ is equal to two, that harmonic
will be dominated by the preference of particles to flow in the
direction of the reaction plane. This feature has been exploited to
study charge separation relative to the reaction plane through
measurements of the charge dependence of
$\langle\cos(\phi_1+\phi_2-2\phi_3)\rangle$~\cite{Abelev:2009ac,Abelev:2009ad}. The
motivation for those measurements was to search for evidence of the
chiral magnetic effect (CME) in heavy-ion
collisions~\cite{Kharzeev:1998kz,Kharzeev:2004ey,Voloshin:2004vk}. By
extending the measurements to other harmonics we can ascertain more
information about the nature of the correlations
interpreted as evidence for CME. Finally, three-particle correlations
reveal information about how various harmonics are correlated with
each other. For example, Teaney and Yan~\cite{Teaney:2010vd}
originally proposed the measurement of
$\langle\cos(\phi_{1}+2\phi_{2}-3\phi_{3})\rangle$ because initial
state models predict a strong correlation between the first, second
and third harmonics of the spatial density distribution. That
correlation can be traced to collision geometries where a nucleon from
one nucleus fluctuates toward the edge of that nucleus and impinges on
the oncoming nucleus. This leads to something similar to a $p+A$
collision and a high density near the edge of the main collision
region. That configuration increases the predicted $v_3$ by a factor
of 2-3 in noncentral collisions so that $v_3$ deviates from the
1/$\sqrt{N_{\mathrm{part}}}$ one would expect from random fluctuations
in the positions of the nucleons participating in the
collision~\cite{Sorensen:2011hm,earlyv3,besv3}. That configuration should also be asymmetric in the forward and backward rapidity directions, again pointing to the importance of understanding the three dimensional structure of the initial state. If the evidence
proposed by Teaney and Yan is not confirmed, then one may question
the validity of any model that predicts the centrality dependence of
$v_n$ based on those initial condition models. In this paper we
present measurements of $\langle\cos(m\phi_1+n\phi_2-(m+n)\phi_3)\rangle$ as
a function of energy, centrality, $\Delta\eta$, $p_T$, and harmonics
$m$ and $n$. Data confirm the predicted correlation between the first, second and third harmonics but the $\Delta\eta$ dependence points to the potential importance of including the three-dimensional structure of the initial state in the model calculations.

In the following, we first describe the experiment and the analysis (Sec.~\ref{an}). We then present the results in Sec.~\ref{re} including the $\Delta\eta$ dependence (Sec.~\ref{eta}), the centrality dependence (Sec.~\ref{cent}), the $p_T$ dependence (Sec.~\ref{pt}), and the beam energy dependence (Sec.~\ref{edepsec}). Conclusions are presented in Sec.~\ref{co}. We include measurements of $v_n^2\{2\}$ for n=1,2,4, and 5 in the appendix.

\section{Experiment and Analysis}
\label{an}

Our measurements make use of data collected from Au+Au collisions with
the STAR detector at RHIC in the years 2004, 2010, 2011, 2012, and 2014.
The charged particles used in this analysis are detected through their
ionization energy loss in the STAR Time Projection
Chamber~\cite{STAR}. The transverse momentum $p_T$, $\eta$, and charge
are determined from the trajectory of the track in STAR's solenoidal
magnetic field. With the 0.5 Tesla field used during data taking,
particles can be reliably tracked for $p_T>0.2$ GeV/$c$. The efficiency
for finding particles drops quickly as $p_T$ decreases below this
value~\cite{Abelev:2008ab}. Weights have been used to correct the
three-particle correlation functions for the $p_T$-dependent
efficiency and for imperfections in the detector acceptance.  The
quantity analyzed and reported is
\begin{multline}
C_{m,n,m+n}= \langle \cos (m\phi_1+n\phi_2-(m+n)\phi_3) \rangle =\\
 \left\langle \left( \frac{\sum_{i,j,k} w_iw_jw_k\cos (m\phi_i+n\phi_j+(m+n)\phi_k)}{\sum_{i,j,k}w_iw_jw_k}\right) \right\rangle
\end{multline}
where $\langle\rangle$ represents an average over events and
$\sum_{i,j,k}$ is a sum over unique particle triplets within an
event. Each event is weighted by the number of unique triplets in that
event. The weights $w_{i,j,k}$ are determined from the inverse of the
$\phi$ distributions after they have been averaged over many events
(which for a perfect detector should be flat) and by the $p_T$
dependent efficiency. The $w_{i,j,k}$ depend on the particles' $p_T$,
$\eta$, and charge and the collisions' centrality and z-vertex
location. The correction procedure is verified by checking that the
$\phi$ distributions are flat after the correction so that
$\langle\cos n(\phi)\rangle$ and $\langle\sin n(\phi)\rangle$ are near zero. With
these corrections, the data represent the {\cijk} that would be seen
by a detector with perfect acceptance for particles with $p_T>0.2$
GeV/$c$ and $|\eta|<1$. In practice, calculating all possible
combinations of three particles individually would be computationally
too costly to be practical, particularly for the larger data sets at
200 GeV. In that case we use algebra based on Q-vectors ($Q_{n}=\Sigma\exp(in\phi)$) to reduce the
computational challenge~\cite{Bilandzic:2010jr}. Differential
measurements like the $\Delta\eta$ dependence of the correlations,
however, require explicit calculations for at least two of the
particles. Studying the $\Delta\eta$ dependence of the correlations
also allows us to correct for the effect of track-merging on the
correlations. Track-merging leads to a large anti-correlation between
particle pairs that are close to each other in the detector. The
effect becomes large in central collisions where the detector
occupancy is largest. After weight corrections have been applied to
correct for single particle acceptance effects, the effect of
track-merging is the largest remaining correction. Data have been
divided into standard centrality classes (0-5\%, 5-10\%,
10-20\%,... 70-80\%) based on the number of charged hadrons within $|\eta|<0.5$ observed
for a given event. In some figures, we will report
the centrality in terms of the number of participating nucleons
({\npart}) estimated from a Monte Carlo Glauber
calculations~\cite{Abelev:2008ab,glauber}.

The three-particle correlations presented in this paper are related
to the low-resolution limit of the event-plane measurements that have been explored at the
LHC~\cite{Aad:2014fla}. Practically this would be carried out by
dividing {\cijk} by $\langle v_mv_nv_{m+n}\rangle$. Typically,
however, $v_{n}$ is measured from a two-particle correlation function
such as the two-particle cumulants $v_n=\sqrt{v_{n}^2\{2\}}$ or a
similar measurement and the $v_{n}^2\{2\}$ are not positive-definite
quantities. As such, $\sqrt{v_{n}^2\{2\}}$ can, and often does, become
imaginary. This is particularly true for the first harmonic and also at
lower collision energies. For this reason we report the pure
three-particle correlations which, in any case, do not suffer from the
ambiguities related to the low- and high-resolution limits associated with reaction plane analyses~\cite{Bhalerao:2013ina,Luzum:2012da} and are therefore
easier to interpret theoretically.

\section{Results}
\label{re}

In the following, we present the $\Delta\eta$ dependence of the
three-particle correlations for several harmonic combinations
corrected for track-merging. After removing the effects of track
merging and Hanbury Brown and Twiss (HBT) correlations~\cite{Lisa:2005dd}, we integrate over the
$\Delta\eta$ dependence of the correlations and present the resulting
integrated correlations as a function of centrality for the energies
$\sqrt{s_{{\rm NN}}}$=200, 62.4, 39, 27, 19.6, 14.5, 11.5, and 7.7 GeV.
We also investigate the $p_T$ dependence of the correlations by
plotting them as a function of the $p_T$ of either the first or second
particle used in the correlation. Finally, we study how the data depends on the beam energy.

\subsection{$\Delta\eta$ Dependence}
\label{eta}

\begin{figure}[htbp]
\centering\mbox{
\includegraphics[width=0.49\textwidth]{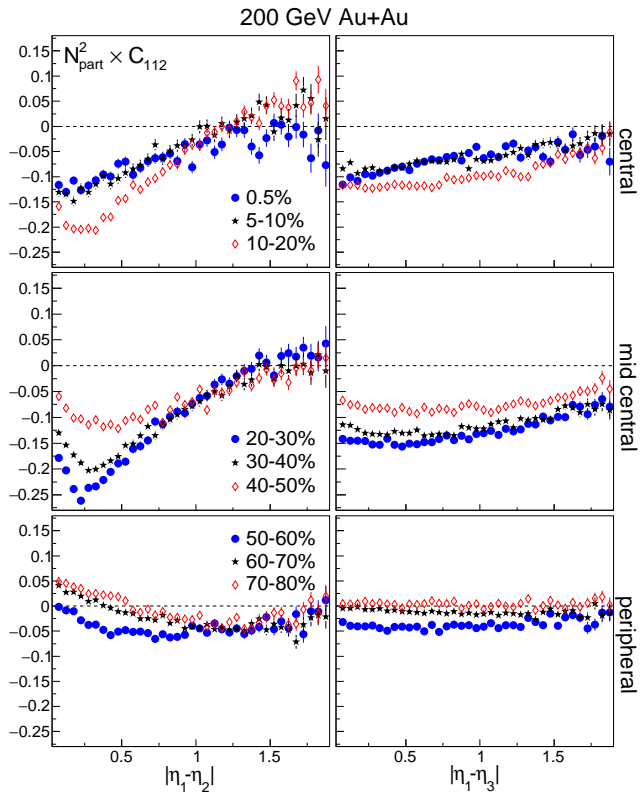}}
\caption{ (color online) The $\Delta\eta$ dependence of {\caab} scaled
  by $N_{\mathrm{part}}^{2}$ for 9 centrality intervals with the three
  most central classes shown in the top panels and the three most
  peripheral in the bottom. The {\npart} values used for
  the corresponding centralities are 350.6, 298.6, 234.3, 167.6,
  117.1, 78.3, 49.3, 28.2 and 15.7. In the panels on the left, $\Delta\eta$ is
  taken between particles 1 and 2 while on the right it is between
  particles 1 and 3 (which is identical to 2 and 3). Data are from 200 GeV
  Au+Au collisions and for charged hadrons with $p_T>0.2$ GeV/$c$,
  $|\eta|<1$.} \label{fig1}
\end{figure}

Figure~\ref{fig1} shows the $\Delta\eta$ dependence of {\caab} scaled by
$N_{\mathrm{part}}^2$ for charged hadrons with $p_T>0.2$ GeV/$c$ and
$|\eta|<1$. The scaling accounts for the natural dilution of
correlations expected if the more central collisions can be treated as a
linear superposition of nucleon-nucleon collisions. Results for nine different
centrality intervals from 200 GeV Au+Au collisions are shown. We do not include the uncertainty on {\npart} in the uncertainties in our figures. The left
panels show the correlations as a function of the difference in $\eta$
between the first and second particle. Note that the subscripts in {\cijk} refer to the harmonic number while the subscripts for the $\eta$ refers to the particle number. The right panels show the same
but as a function of the difference between particles 1 and 3. The
{\caab} correlation is similar to the correlation used in the search
for the chiral magnetic effect except that we do not separate out the
cases when particles 1 and 2 have like-sign charges vs unlike-sign
charges as is done when looking for charge separation with respect to
the reaction plane.  These measurements can be approximately related to
the reaction-plane based measurements by scaling the three-particle
correlations by 1/$v_2$. We note that the difference in {\caab} for different charge combinations is as large as the signal with {\caab} being nearly zero for unlike-sign combitions of particle 1 and 2. This correlation may also be influenced by momentum conservation effects as well. It's not clear however how those effects would be distributed with respect to $\Delta\eta$.

In the left panels of Fig.~\ref{fig1}, we see a strong dependence for {\caab} on
$|\eta_1-\eta_2|$. In central collisions, the data starts out negative
at the smallest values of $|\eta_1-\eta_2|$ but then begins to
increase and becomes close to zero or even positive near
$|\eta_1-\eta_2|=1.5$. At small $|\eta_1-\eta_2|$, a narrow peak is
seen in the correlation that is related to HBT. As we progress from
central to peripheral collisions, the trends change with {\caab} in
peripheral collisions exhibiting a positive value at small
$|\eta_1-\eta_2|$, perhaps signaling the dominance of jets in the
correlation function in these peripheral collisions.

The left panels share the same scales as the right panels making it
clear that the dependence of {\caab} on $|\eta_1-\eta_3|$ is much
weaker than the dependence on $|\eta_1-\eta_2|$. This is expected
since the $e^{-2i\phi_3}$ term in {\caab}=$\langle
e^{i\phi_1}e^{i\phi_2}e^{-2i\phi_3}\rangle$ will be dominated by the
global preference of particles to be emitted in the direction of the
reaction plane. For all but the most central collisions, the almond shaped geometry of the collision overlap region is approximately invariant with rapidity. This is not likely the case for other harmonics.

\begin{figure}[htbp]
\centering\mbox{
\includegraphics[width=0.49\textwidth]{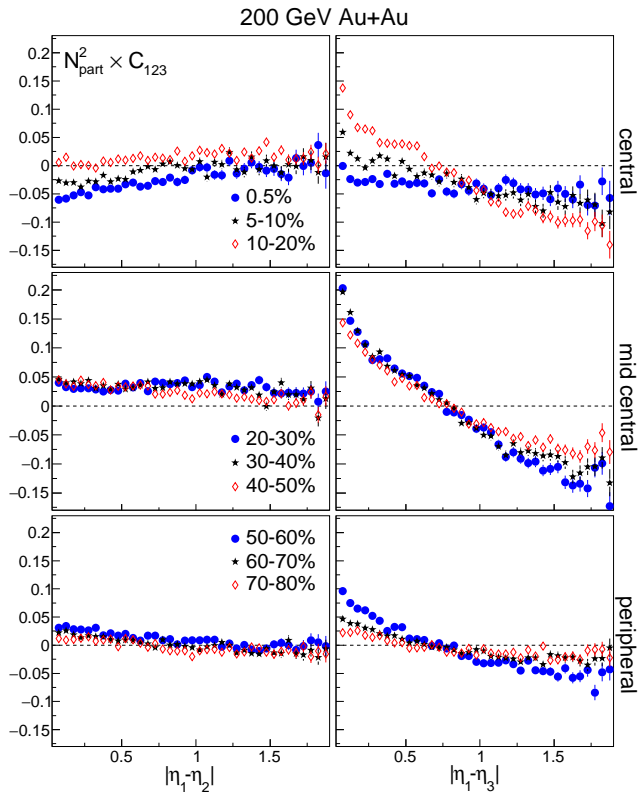}}
\caption{ (color online). The $\Delta\eta$ dependence of {\cabc}
  scaled by $N_{\mathrm{part}}^2$ for 9 centrality intervals with the
  three most central classes shown in the top panels and the three
  most peripheral in the bottom. In the panels on the left,
  $\Delta\eta$ is taken between particles 1 and 2 while on the right
  it is between particles 1 and 3. Data are from 200 GeV Au+Au
  collisions and for charged hadrons with $p_T>0.2$ GeV/$c$,
  $|\eta|<1$.} \label{fig2}
\end{figure}

Figure~\ref{fig2} shows {\cabc} scaled by $N_{\mathrm{part}}^2$ as a
function of $|\eta_1-\eta_2|$ (left panels) and $|\eta_1-\eta_3|$
(right panels). In this case, {\cabc} exhibits a stronger dependence
on $|\eta_1-\eta_3|$ than on $|\eta_1-\eta_2|$. The variation with
$|\eta_2-\eta_3|$ is very similar to the variation with
$|\eta_1-\eta_2|$ and is omitted from the figures to improve legibility. Again,
the $e^{i2\phi_2}$ component of {\cabc} is dominated by the reaction
plane which is largely invariant within the $\eta$ range covered by
these measurements so that {\cabc} depends very little on the
$\eta_2$, $|\eta_1-\eta_2|$, or $|\eta_2-\eta_3|$. However, {\cabc}
depends very strongly on $|\eta_1-\eta_3|$. This dependence may arise
from the longitudinal asymmetry inherent in the fluctuations that lead
to predictions for large values of {\cabc}~\cite{Teaney:2013dta}. In
models for the initial geometry, the correlations are induced between
the first, second, and third harmonics of the eccentricity by cases
where a nucleon fluctuates towards the edge of the
nucleus~\cite{Shou:2014eya}. If that occurs in the reaction plane
direction and towards the other nucleus in the collision, then that
nucleon can collide with many nucleons from the other nucleus. This geometry will cause the first and third harmonics to become correlated with the second harmonic. Since the collision of one nucleon from one nucleus with
many nucleons in the other nucleus is asymmetric along the rapidity axis, we argue that we
can expect a strong dependence on $|\eta_1-\eta_3|$. Models that
assume the initial energy density is symmetric with rapidity (boost invariant) will
likely fail to describe this behavior. One may also speculate that the
variation with $|\eta_1-\eta_3|$ could arise from sources like jets or
resonances particularly if they interact with the medium so that they become correlated with the reaction plane. Making use of the full suite of measurements provided here will help delineate between these two scenarios.

\begin{figure}[htbp]
\centering\mbox{
\includegraphics[width=0.49\textwidth]{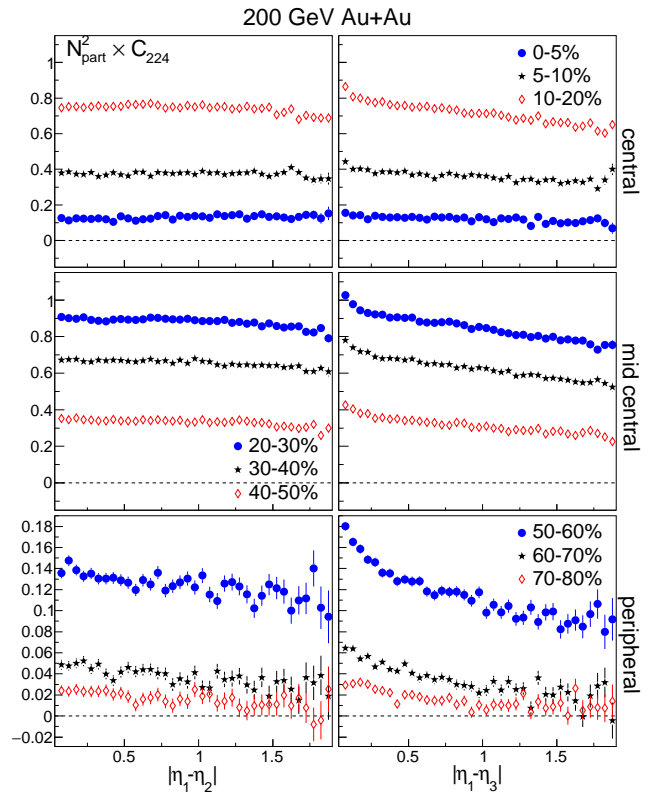}}
\caption{ (color online)  The $\Delta\eta$ dependence of {\cbbd} scaled by $N_{\mathrm{part}}^2$ for
  9 centrality intervals with the three most central classes shown in
  the top panels and the three most peripheral in the bottom. In the
  panels on the left, $\Delta\eta$ is taken between particles 1 and 2
  while on the right it is between particles 1 and 3 (identical to 2
  and 3). Data are from 200 GeV Au+Au collisions and for charged
  hadrons with $p_T>0.2$ GeV/$c$, $|\eta|<1$. } \label{fig3}
\end{figure}

In Fig.~\ref{fig3} we present the $|\eta_1-\eta_2|$ and
$|\eta_1-\eta_3|$ dependence of {\cbbd}. This correlation is more
strongly influenced by the reaction plane correlations and exhibits
much larger values than either {\caab} or {\cabc}. The dependence on
$|\eta_1-\eta_2|$ and $|\eta_1-\eta_3|$ are also weaker with {\cbbd} in central and mid-central collisions
showing little variation over the $|\eta_1-\eta_2|$ range, consistent with
a mostly $\eta$-independent reaction plane within the measured range. A larger variation is observed with $|\eta_1-\eta_3|$ which in mid-central collisions amounts to an approximately 20\% variation. We also note that in mid-central collisions, the change in value of {\cbbd} over the range $0<|\eta_1-\eta_3|<2$ is similar in magnitude to the change of {\caab} over $0<|\eta_1-\eta_2|<2$ and {\cabc} over $0<|\eta_1-\eta_3|<2$.

\begin{figure}[htbp]
\centering\mbox{
\includegraphics[width=0.49\textwidth]{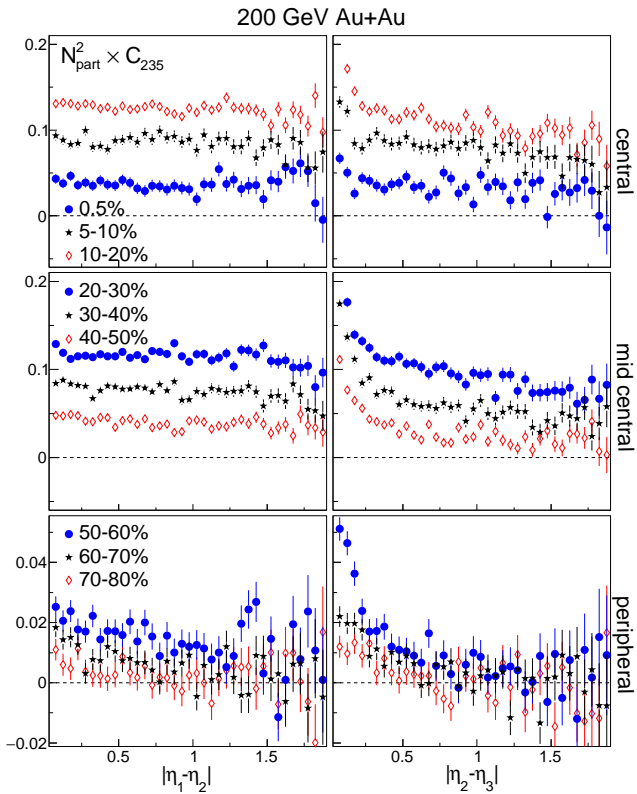}}
\caption{ (color online) The $\Delta\eta$ dependence of {\cbce} scaled
  by $N_{\mathrm{part}}^2$ for 9 centrality intervals with the three
  most central classes shown in the top panels and the three most
  peripheral in the bottom. In the panels on the left, $\Delta\eta$ is
  taken between particles 1 and 2 while on the right it is between
  particles 1 and 3 (identical to 2 and 3). Data are from 200 GeV
  Au+Au collisions and for charged hadrons with $p_T>0.2$ GeV/$c$,
  $|\eta|<1$.} \label{fig4}
\end{figure}

In Fig.~\ref{fig4} we present the $|\eta_1-\eta_2|$ and
$|\eta_2-\eta_3|$ dependence of {\cbce}. Again, {\cbce} only exhibits a
weak dependence on $|\eta_1-\eta_2|$ but a stronger dependence on
$|\eta_2-\eta_3|$. In central and mid-central collisions, a strong short-range correlation at
$|\eta_2-\eta_3|<0.4$ is apparent consistent with HBT and Coulomb
correlations that vary with respect to the reaction plane. In addition to that peak, {\cbce} decreases as $|\eta_2-\eta_3|$ increases. Although the relative variation of ${\cbce}$ is similar to {\cbbd}, the absolute change is much smaller than for {\caab}, {\cabc}, or {\cbbd}. 

The combination of the various {\cijk} can help elucidate the nature
of the three-particle correlations. If the $|\eta_1-\eta_3|$ dependence of
{\cabc} arises from correlations between particles from jets correlated with the reaction plane, we would
expect the particles at small $\Delta\eta$ to predominantly come from the near-side
jet (at $\Delta\phi \approx 0$) and particles at larger $\Delta\eta$ to come from the away-side
jet (at $\Delta\phi \approx \pi$ radians). In that case, at small $\Delta\eta$, {\cijk} for all harmonics will
have a positive contribution from the jets. The same
is not true however for large $\Delta\eta$ where we would expect the
correlations to be dominated by the away-side jet separated by $\pi$ radians. For
this case at large $\Delta\eta$, {\caab} and {\cabc} would receive negative contributions from the away
side jet
while {\cbbd} and {\cbce} would both receive positive
contributions. 
The trends observed across the variety of {\cijk} measurements
are inconsistent with this simple picture with {\cbbd} decreasing by nearly the same amount as {\cabc} as $\Delta\eta$ is increased. A more complicated picture of the effect of jets would therefore be required to account for the observed data but it appears difficult to construct a non-flow scenario that can account for the long-range variation of {\cijk}. Breaking of boost-invariance in the initial density distributions may provide an explanation for the observed variations but we do not know of any specific model that has been shown to describe our data.

\subsection{Centrality Dependence}
\label{cent}

\begin{figure*}[hbtp]
\centering\mbox{
\includegraphics[width=0.99\textwidth]{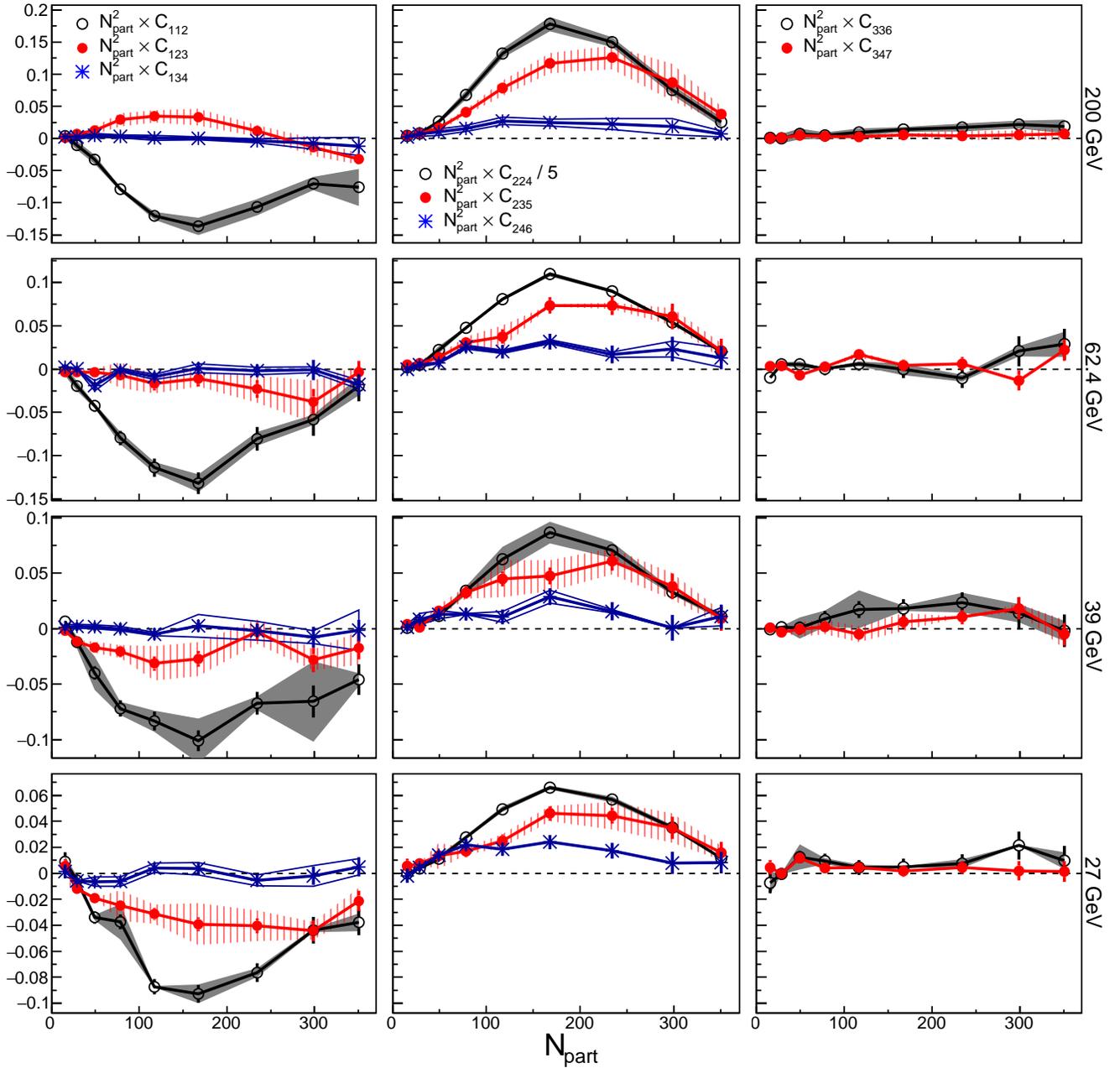}}
\caption{ (color online) The centrality dependence of the {\cijk}
  correlations scaled by $N_{\mathrm{part}}^2$ for charged hadrons with $p_T>0.2$ GeV/$c$ and $|\eta|<1$
  from 200, 62.4, 39, and 27 GeV Au+Au collisions for $(m,n) = (1,1),
  (1,2), (1,3)$ (left) $(2,2), (2,3), (2,4)$ (center) and $(3,3),
  (3,4)$ (right). Systematic errors are shown as bands. All panels in the same row share the same scale but
  {\cbbd} has been divided by a factor of 5 to fit on the
  panel. The labels in the top panels apply to all the panels in same
  column.} \label{cosijk200}
\end{figure*}

\begin{figure*}[hbtp]
\centering\mbox{
\includegraphics[width=0.99\textwidth]{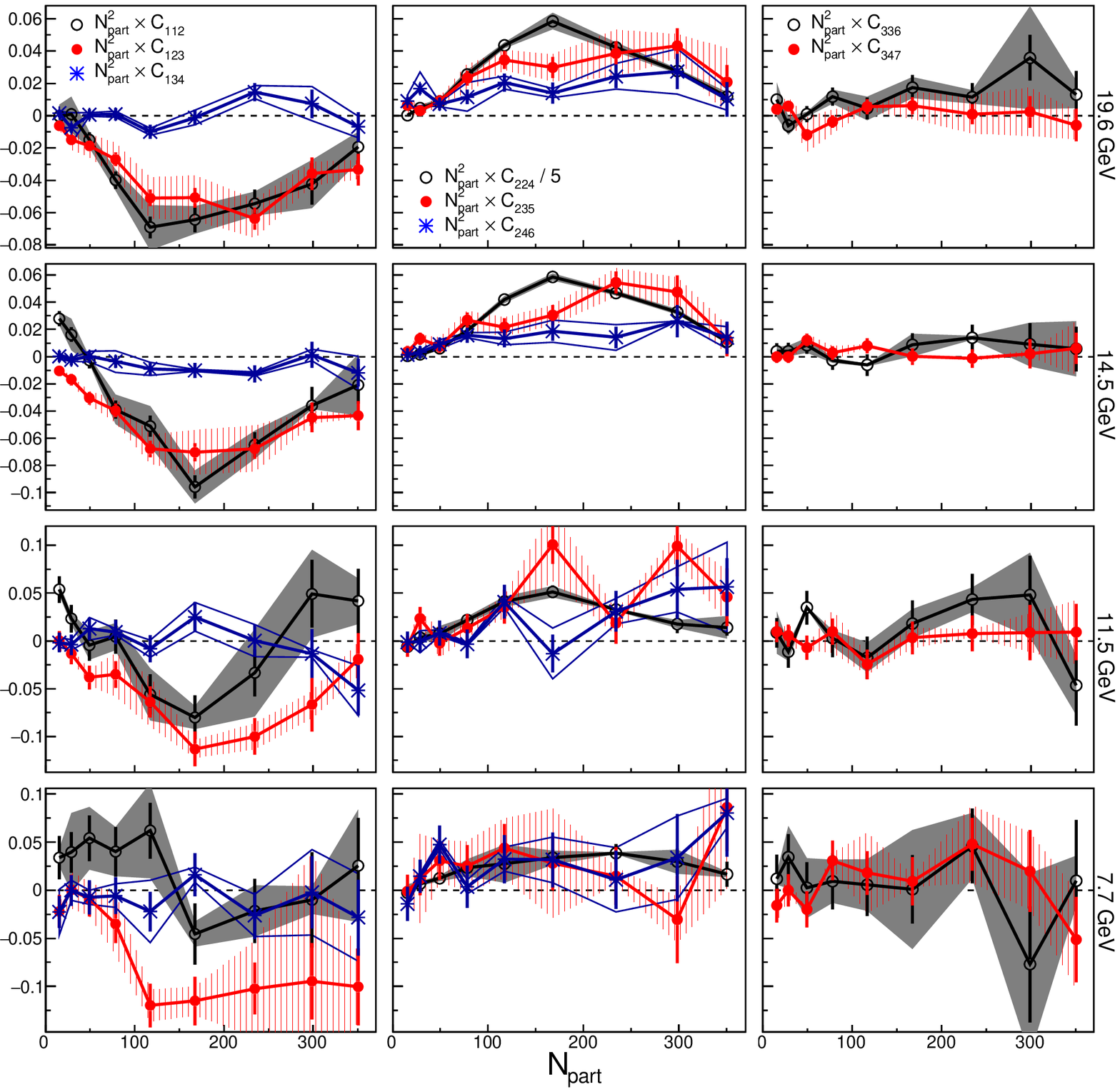}}
\caption{ (color online) The same quantities as Fig.~\ref{cosijk200}
  but for the lower energy Au+Au collisions 19.6, 14.5, 11.5, and 7.7
  GeV.  } \label{cosijk62}
\end{figure*}

In Figs.~\ref{cosijk200} and \ref{cosijk62} we show {\cijk}
correlations scaled by $N_{\mathrm{part}}^2$ with $(m,n)$ = $(1,1)$, $(1,2)$, $(1,3)$, $(2,2)$, $(2,3)$,
$(2,4)$, $(3,3)$, and $(3,4)$ for {\snn}=200, 62.4, 39, 27, 19.6,
14.5, 11.5, and 7.7 GeV Au+Au collisions as a function of
{\npart}. Data are for charged particles with $|\eta|<1$ and
$p_T>0.2$~GeV/$c$.  
The correlation {\cbbd}, by far the largest of the measured
correlations, has been scaled by a factor of 1/5. Otherwise, the
scales on each of the three panels are kept the same for each energy
to make it easier to compare the magnitudes of the different harmonic
combinations.

At 200 GeV, {\caab} is negative for all centralities except for the
most peripheral where it is slightly positive but consistent with
zero. {\cabc} is consistent with zero in peripheral collisions, positive in mid-central collisions but then becomes negative in central collisions. If the second and third harmonic event planes are
uncorrelated, then {\cabc} should be zero. The {\cabc} correlation is
non-zero deviating from that expectation. The magnitude is however
much smaller than originally anticipated based on a linear
hydrodynamic response to initial state geometry
fluctuations~\cite{Teaney:2010vd}. Non-linear coupling between
harmonics, where the fifth harmonic for example is dominated by a combination of the second and third harmonic, has been shown to be very
important~\cite{Qiu:2012uy,Teaney:2012ke}. In the case of {\cabc}, the
non-linear contribution has an opposite sign to the linear
contribution and similar magnitude canceling out most of the expected
strength of {\cabc}. This suggests that {\cabc} is very sensitive to
the nonlinear nature of the hydrodynamic model. {\cacd} is close to zero for all centralities indicating little or no correlation between the first, third, and fourth harmonics. The other {\cijk} correlations are positive for all
centralities.  When considering the comparison of this data to hydrodynamic models, it is important to also consider the strong $\Delta\eta$ dependence of the correlations as shown in the previous section.

The correlations involving a second harmonic are largest with {\cbbd}
being approximately 5 times larger in magnitude than the next largest
correlator {\cbce}. The correlations decrease quickly as harmonics are
increased beyond n=2. The higher harmonic correlations {\cccf} and
{\ccdg} are both small but non-zero. The correlations {\caab},
{\cabc}, {\cbbd}, {\cbce}, and {\cccf} scaled by {\npartsq} all exhibit extrema in mid
central collisions where the initial overlap geometry is predominantly
elliptical. We note that the centrality at which {\npartsq}{\cbbd} reaches a maximum is different than the centrality at which {\npartsq}{\cbce} reaches a maximum.

As the collision energy is reduced, although the magnitude of the correlations
becomes smaller, the centrality dependence and ordering of the
different harmonics seems to remain mostly the same. The {\cabc}
correlation however is an exception. While at 200 GeV, {\cabc} is
mostly positive, at 62.4 GeV it is consistent with zero or slightly negative and
at lower energies it becomes more and more negative. We speculate that
this behavior may be related to the increasing importance of momentum
conservation as the number of particles produced in the collision
decreases. No theoretical guidance exists however for the energy
dependence of these correlations at energies below 200 GeV. This data should provide useful constraints for the models being developed to describe lower energy collisions associated with the energy scan program at RHIC.

Figure~\ref{cosijk62} shows the same correlations as Fig.~\ref{cosijk200}
except for lower energy data sets: $\sqrt{s_{_{{\rm NN}}}}=$ 19.6, 14.5, 11.5, and 7.7 GeV. Trends similar to those seen in Fig.~\ref{cosijk200} are for the most part also exhibited in this
figure. Although the statistical precision is poor for the lowest
energy points, it appears that {\caab} at 7.7 GeV is smaller in magnitude than at higher energies, becoming consistent with zero. This was also observed in the
charge dependent measurements of {\caab}~\cite{Adamczyk:2014mzf}. A second phase of the RHIC
beam energy scan planned for 2019 and 2020 will significantly increase
the number of events available for analysis at these lower energies while expanding the $\eta$ acceptance from $|\eta|<1$ to $|\eta|<1.5$~\cite{itpc} so
that this intriguing observation can be further investigated. The increased acceptance will increase the number of three-particle combinations by approximately a factor of three and will make it possible to measure the $\Delta\eta$ dependence of the {\cijk} correlations to $|\Delta\eta| \approx 3$.

\subsection{$p_{T}$ Dependence}
\label{pt}

\begin{figure}[hbtp]
\centering\mbox{
\includegraphics[width=0.49\textwidth]{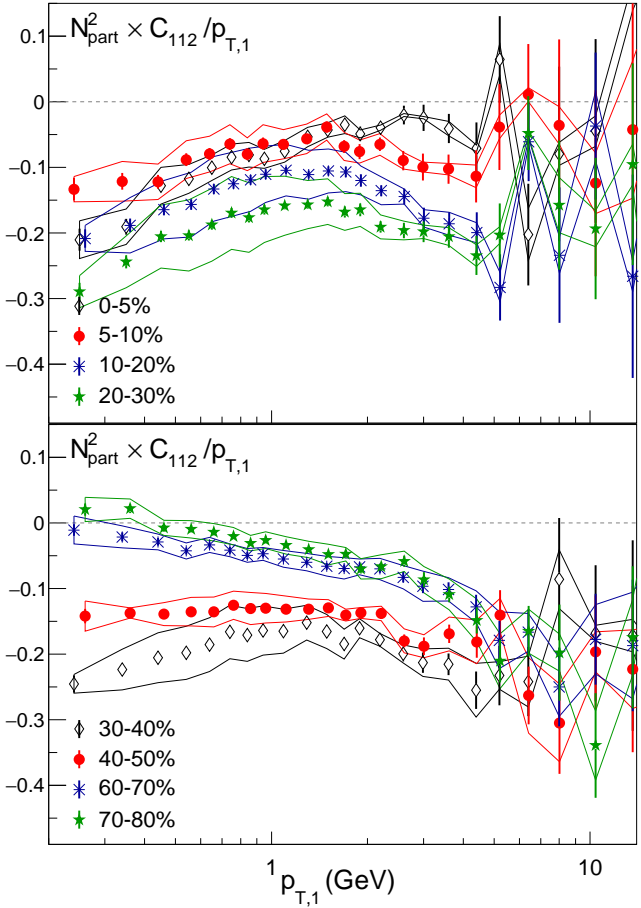}}
\caption{ (color online) Three-particle azimuthal correlations {\caab} scaled by {\npartsq}/$p_{T,1}$ as a function of the first particles $p_T$ for 200 GeV Au+Au
  collisions for charged hadrons with $p_T>0.2$ GeV/$c$ and
  $|\eta|<1$. The top and bottom panels show the same quantity but for
  a different set of centrality intervals. Systematic errors are shown
  as solid lines enclosing the respective data
  points.} \label{pt112}
\end{figure}

If the three-particle correlations presented here are dominated by
correlations between event planes, then one might expect that the
$p_T$ dependence of the three-particle correlations will simply track
the $p_T$ dependence of the relevant $v_n$~\cite{Teaney:2010vd}: 
\begin{multline}
\langle \cos(m\phi_1(p_T)+n\phi_2-(m+n)\phi_3)\rangle 
\approx \\
\frac{v_m(p_T)}{\varepsilon_m}\frac{v_n}{\varepsilon_n}\frac{v_{m+n}}{\varepsilon_{m+n}}\times \\
\langle \varepsilon_m\varepsilon_{n}\varepsilon_{m+n}\cos(m\Psi_m+n\Psi_n-(m+n)\Psi_{m+n})\rangle,
\label{eqn2}
\end{multline}
where $\varepsilon_m$ is the $m^{{\mathrm{th}}}$ harmonic eccentricity
and $\Psi_{m}$ is the $m^{{\mathrm{th}}}$ harmonic participant plane
angle. For the purpose of simplicity in this publication, we
have scaled the correlations by $N^{2}_{\mathrm{part}}/p_T$ to account for
the general increase of $v_n(p_T)$ with $p_T$~\cite{v2papers}. That
simple scaling is only valid at lower $p_T$ and for $n \neq 1$. It does, however, aid in visualizing trends in the data which would otherwise be visually dominated by the larger $p_T$ range. Our primary reason for introducing Eq.~\ref{eqn2} is to provide a context for understanding the $p_T$ dependence of {\cijk}. The relationship between {\cijk} and harmonic planes in Eq.~\ref{eqn2} is not guaranteed to hold and is particularly likely to be broken for correlations involving the first harmonic where momentum conservation effects will likely play an important role or where a strong charge sign dependence has been observed~\cite{Abelev:2009ac,Abelev:2009ad}.

\begin{figure*}[hbtp]
\centering\mbox{
\includegraphics[width=0.89\textwidth]{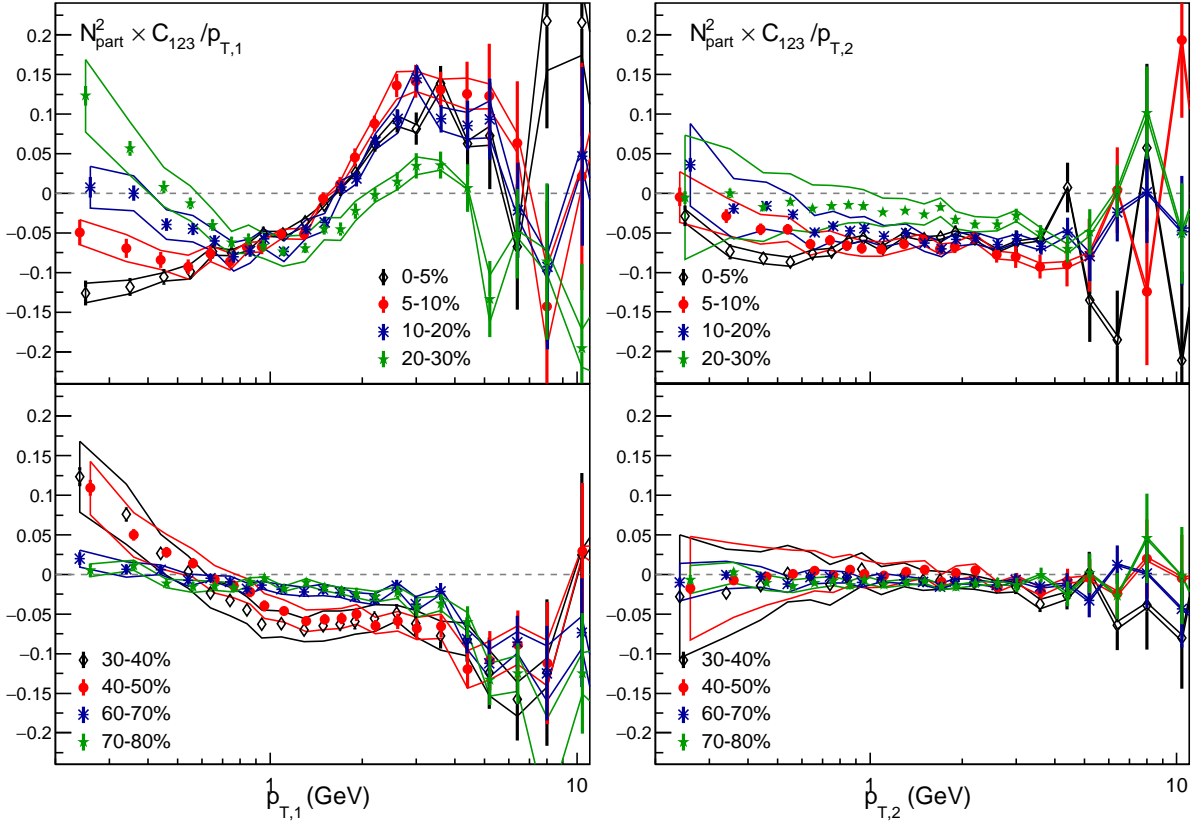}}
\caption{ (color online) Three-particle azimuthal correlations {\cabc} scaled by {\npartsq}/$p_T$
  as a function of the $p_T$ using the $p_T$ of particle one (left panels) or of particle
  two (right panels) for 200 GeV Au+Au collisions. Data are for
  charged hadrons with $p_T>0.2$ GeV/$c$ and $|\eta|<1$. The top and
  bottom panels show the same quantity but for a different set of
  centrality intervals. Systematic errors are shown as solid lines
  enclosing the respective data points.  } \label{pt123}
\end{figure*}

In Fig.~\ref{pt112} we show {\npartsq}{\caab}/$p_T$ as a function of
the $p_T$ of particle one. The top panel shows the more central
collisions while the bottom panel shows more peripheral collisions. In this and in
the following figures related to the $p_T$ dependence, we sometimes
exclude centrality bins and slightly shift the positions
of the points along the $p_T$ axis to make the figures more
readable. For more central collisions, {\caab}/$p_{T,1}$ is negative and
slowly decreases in magnitude as $p_{T,1}$ increases. This indicates that
{\caab} is generally increasing with the $p_T$ of particle one but
that for central collisions at high $p_T$, {\caab} starts to
saturate. For the more peripheral 30-40\% and 40-50\% collision
however, {\caab} appears to be linear in $p_T$ without an indication
of saturation even up to $p_T \approx 10$~GeV/$c$. For the much
more peripheral 60-70\% and 70-80\% centrality intervals, {\caab}
starts out at or above zero then becomes more and more negative as
$p_T$ is increased. The trends in the most
peripheral centrality intervals, particularly at high $p_T$, are consistent with being
dominated by momentum conservation and jets. A pair of back-to-back
particles aligned with the reaction plane will lead to a negative value
for ${\caab}$. Although the data exhibit a smooth transition from the
trends in more central collisions to the trends in more peripheral
collisions, the trends are quite distinct and indicative of very
different correlations in those different regions. In peripheral collisions, the correlations get stronger as $p_T$ is increased. In central collisions, the opposite is observed.

For the case of {\cabc} in Fig.~\ref{pt123}, we show the $p_T$
dependence of both particle one (left panels) and particle two (right
panels). The dependence of {\cabc}$/p_{T,2}$ on $p_{T,2}$
is quite weak indicating that where {\cabc} is non-zero, it increases
roughly linearly with $p_{T,2}$. The dependence of {\cabc}/$p_{T,1}$ on
$p_{T,1}$, however, exhibits several notable trends. First we
note that for the 20-30\% centrality interval, {\cabc}$/p_{T,1}$ changes
sign up to three times. In hydrodynamic models, the value of {\cabc}
is very sensitive to the interplay between linear and non-linear
effects and to viscous effects. The sign
oscillations exhibited in the data may be a consequence of subtle
changes in the relevant sizes of those effects. If this is the case,
then this confirms that {\cabc} is a powerful measurement to help tune
those models. At intermediate $p_{T,1}$ (2-5 GeV/$c$), {\cabc} is positive
for central collisions but negative for peripheral collisions. At
$p_T>7$ GeV/$c$, {\cabc} is strongly negative, perhaps again, indicative
of the contribution of back-to-back jets to the correlations. Such
strong negative correlation seems to be absent in central collisions
where {\cabc} appears to remain positive, although with large error
bars. This is consistent with a scenario where di-jets have been quenched in central collisions. As with {\caab}, the $p_T$ trends for {\cabc} are very different in the most peripheral and most central collisions.

\begin{figure}[hbtp]
\centering\mbox{
\includegraphics[width=0.44\textwidth]{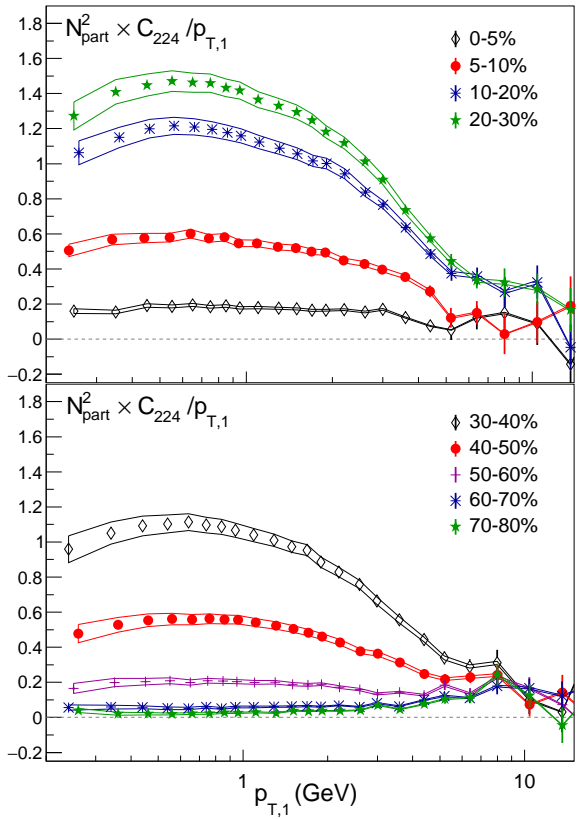}}
\caption{ (color online) Three-particle azimuthal correlations {\cbbd} scaled by {\npartsq}/$p_{T,1}$
  as a function of $p_{T,1}$ for 200 GeV Au+Au
  collisions. Data are for charged hadrons with $p_T>0.2$ GeV/$c$ and
  $|\eta|<1$. The top and bottom panels show the same quantity but for
  a different set of centrality intervals. Systematic errors are shown
  as solid lines enclosing the respective data points.
} \label{pt224}
\end{figure}

The {\cbbd} correlation is the largest of the {\cijk} correlations
since it is strongly affected by the tendency of particles to
preferentially line up with the reaction plane. In Fig.~\ref{pt224}
we show {\npartsq}{\cbbd}$/p_{T,1}$ as a function of $p_{T,1}$. At low $p_{T,1}$, the centrality dependence of the correlations is
as expected from Fig.~\ref{cosijk200} (top panels) where we saw that the integrated
value of {\npartsq}{\cbbd} is largest for mid-central collisions. This
is a natural consequence of the fact that the initial second harmonic
eccentricity decreases as collisions become more central while the
efficiency of converting that eccentricity into momentum-space correlations
increases (with multiplicity). The competition of these two trends leads to a maximum for
second harmonic correlations in mid-central collisions. This
well-known~\cite{v2papers} and generic trend does not persist to
higher values of $p_{T,1}$. We see a clear change in trends at $p_{T,1}>5$ GeV/$c$ with
the most peripheral collisions having the largest correlation strength
while {\npartsq}{\cbbd}$/p_{T,1}$ drops significantly as a function of $p_{T,1}$
for the mid-central collisions. We note that past measurements of
$p_T$ spectra and $v_2(p_T)$ for identified particles have indicated
that the effects of flow may persist up to 5 or 6
GeV/$c$~\cite{v2papers}. This observation is consistent with model
calculations that show in a parton cascade even up to $p_T \approx
5$ GeV/$c$ there are a significant number of partons whose final
momentum has been increased by interactions with the
medium~\cite{Molnar:2005hb}. The $p_{T,1}$ dependence of {\cbbd}/$p_{T,1}$ supports that picture as well.

\begin{figure*}[hbtp]
\centering\mbox{
\includegraphics[width=0.89\textwidth]{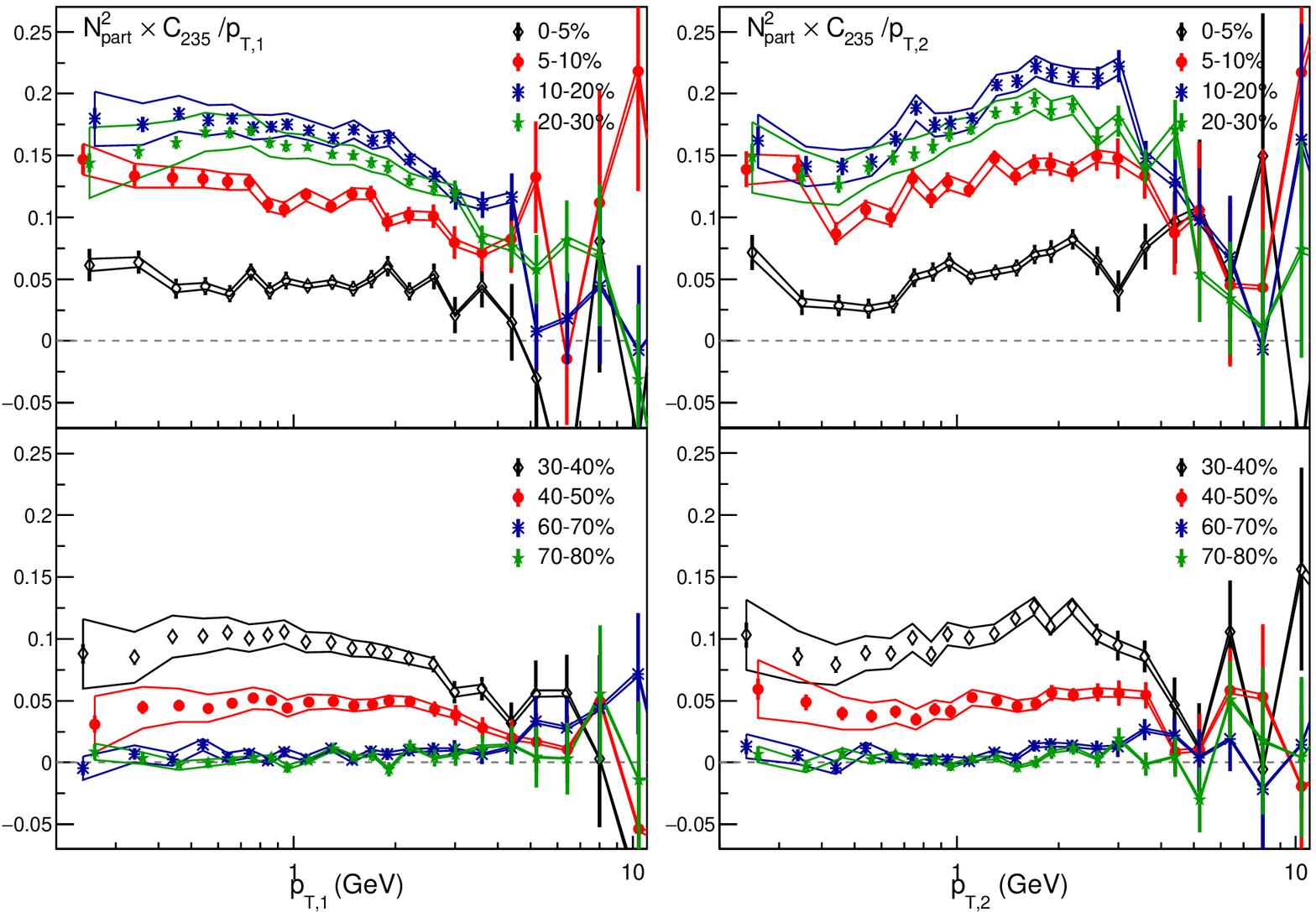}}
\caption{ (color online) Three-particle azimuthal correlations {\cbce} scaled by {\npartsq}/$p_T$
  as a function of $p_T$ where the $p_T$ is taken for either particle one (left panels) or particle
  two (right panels) for 200 GeV Au+Au collisions. Data are for
  charged hadrons with $p_T>0.2$ GeV/$c$ and $|\eta|<1$. The top and
  bottom panels show the same quantity but for a different set of
  centrality intervals. Systematic errors are shown as solid lines
  enclosing the respective data points.  } \label{pt235}
\end{figure*}

In Fig.~\ref{pt235}, we show the $p_{T}$ dependence of {\npartsq}{\cbce}/$p_{T}$ where $p_T$ is either the $p_T$ of 
particle one (left panels) or particle two (right panels). Again, the
top panels show more central collisions and the bottom panels more
peripheral. For $p_T<5$, {\cbce}$/p_T$ is mostly flat as a function of
the $p_T$ of either particle one or particle two. Above that, the
correlations seem to become smaller but with large statistical
errors. One can discern a slight difference between the trends in the left and right panels: {\cbce}/$p_{T,1}$ seems to decrease slightly as a function of $p_{T,1}$, while {\cbce}/$p_{T,2}$ as a function of $p_{T,2}$ seems to increase slightly. This is likely related to the different $p_T$ dependences of $v_2$ and $v_3$ where $v_2$ has been found to saturate at lower $p_T$ while $v_3$ is still growing. In central collisions, it is even found that $v_3$ becomes larger than $v_2$ at intermediate $p_T$~\cite{earlyv3}.

We have tried to point out interesting features in
the $p_T$ dependence of the correlations. In particular, we note that the $p_T$ trends are very different when comparing central collisions to peripheral collisions. We expect that when these
data are compared to model calculations, they will provide even
greater insights into the interplay between the effects of hard
scattering, shear viscosity, bulk viscosity, the collision life-time
and non-linear couplings between harmonics.

\subsection{Energy Dependence}
\label{edepsec}

While Figs.~\ref{cosijk200} and~\ref{cosijk62} show the centrality dependence
of 8 different {\cijk} correlations for 8 beam energies, in this
section we will investigate the energy dependence in greater detail by
first showing the centrality dependence of individual {\cijk}
correlations for a variety of energies in single panels for easier
comparison. We will then show correlations at specific centrality
intervals as a function of {\snn} scaled by $v_2$. Finally we will discuss
implications of the energy dependence of the correlations.

\begin{figure*}[!hbtp]
\centering
\begin{minipage}{0.485\textwidth}
\centering
\includegraphics[width=1.0\textwidth]{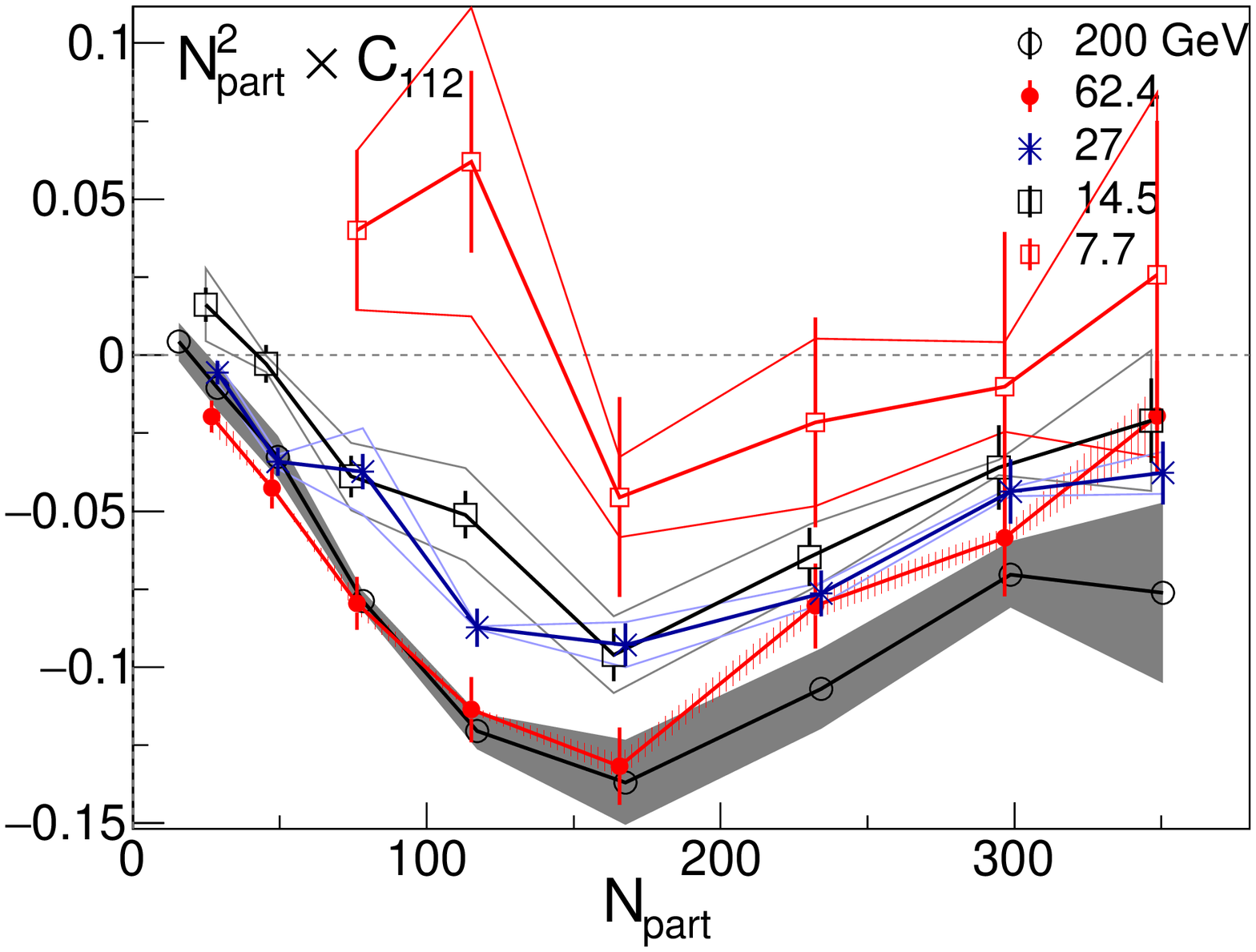}
\end{minipage}
\begin{minipage}{0.485\textwidth}
\centering
\includegraphics[width=1.0\textwidth]{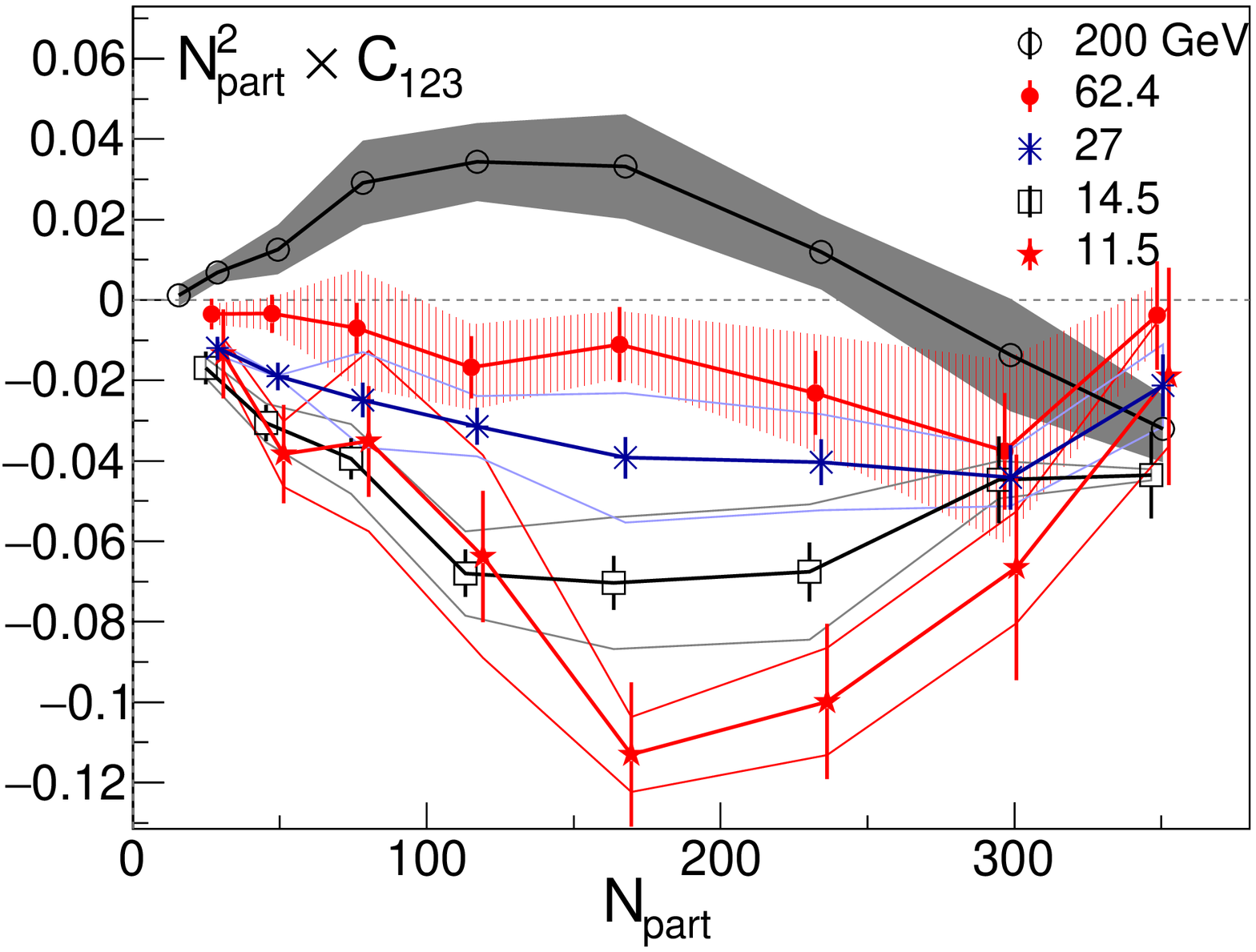}
\end{minipage}
\caption{ (color online) The centrality dependence of {\caab} (left)
  and {\cabc} (right) scaled by {\npartsq} for a selection of energies.  } \label{figa}
\end{figure*}

Figure~\ref{figa} shows the centrality dependence of {\npartsq}{\caab}
(left) and {\npartsq}{\cabc} (right) for 200, 62.4, 27, 14.5, and 7.7 or 11.5
GeV collisions. Some energies are omitted for clarity. For {\npartsq}{\caab},
the general centrality trend appears to remain the same at all
energies except 7.7 GeV, even though the magnitude slightly decreases. For mid-central
collisions, {\caab} is negative for all the energies
shown. The 7.7 GeV data may deviate from the trend observed for the other energeis as will be
discussed later. For {\npartsq}{\cabc}, the energy dependence is quite
different. The only positive values for {\cabc} are for 200 GeV
collisions. At 62.4 GeV, {\npartsq}{\cabc} has a slightly negative value
that is within errors, independent of centrality. As the energy
decreases, {\cabc} becomes more negative so that the centrality
dependence of {\cabc} at 14.5 GeV is nearly the mirror reflection of
the 200 GeV data. As will be discussed below, the change in sign of
{\cabc} has interesting implications for how two-particle correlations relative to the reaction plane change as a function of beam energy.

\begin{figure*}[!hbtp]
\centering
\begin{minipage}{0.485\textwidth}
\centering
\includegraphics[width=1.0\textwidth]{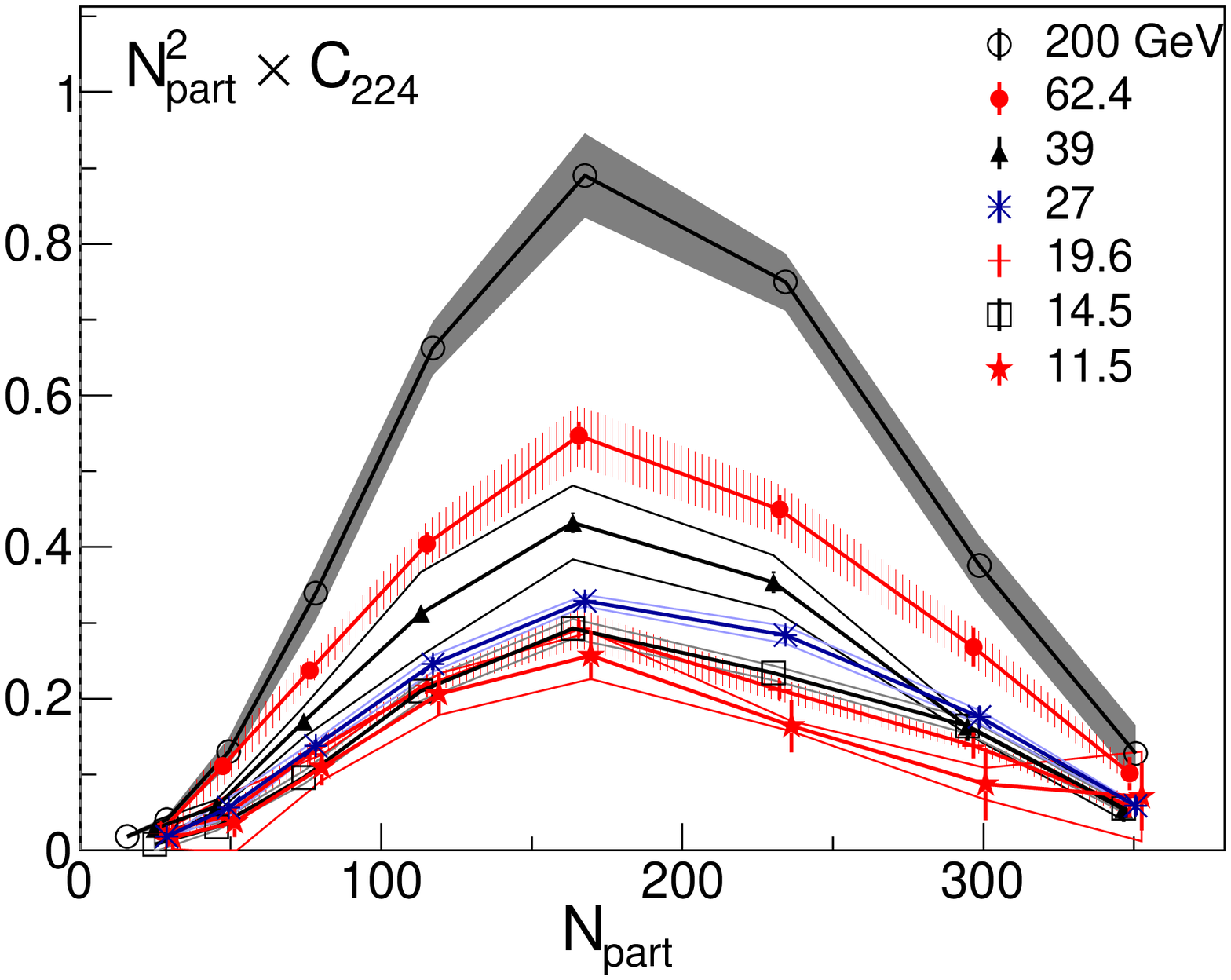}
\end{minipage}
\begin{minipage}{0.485\textwidth}
\centering
\includegraphics[width=1.0\textwidth]{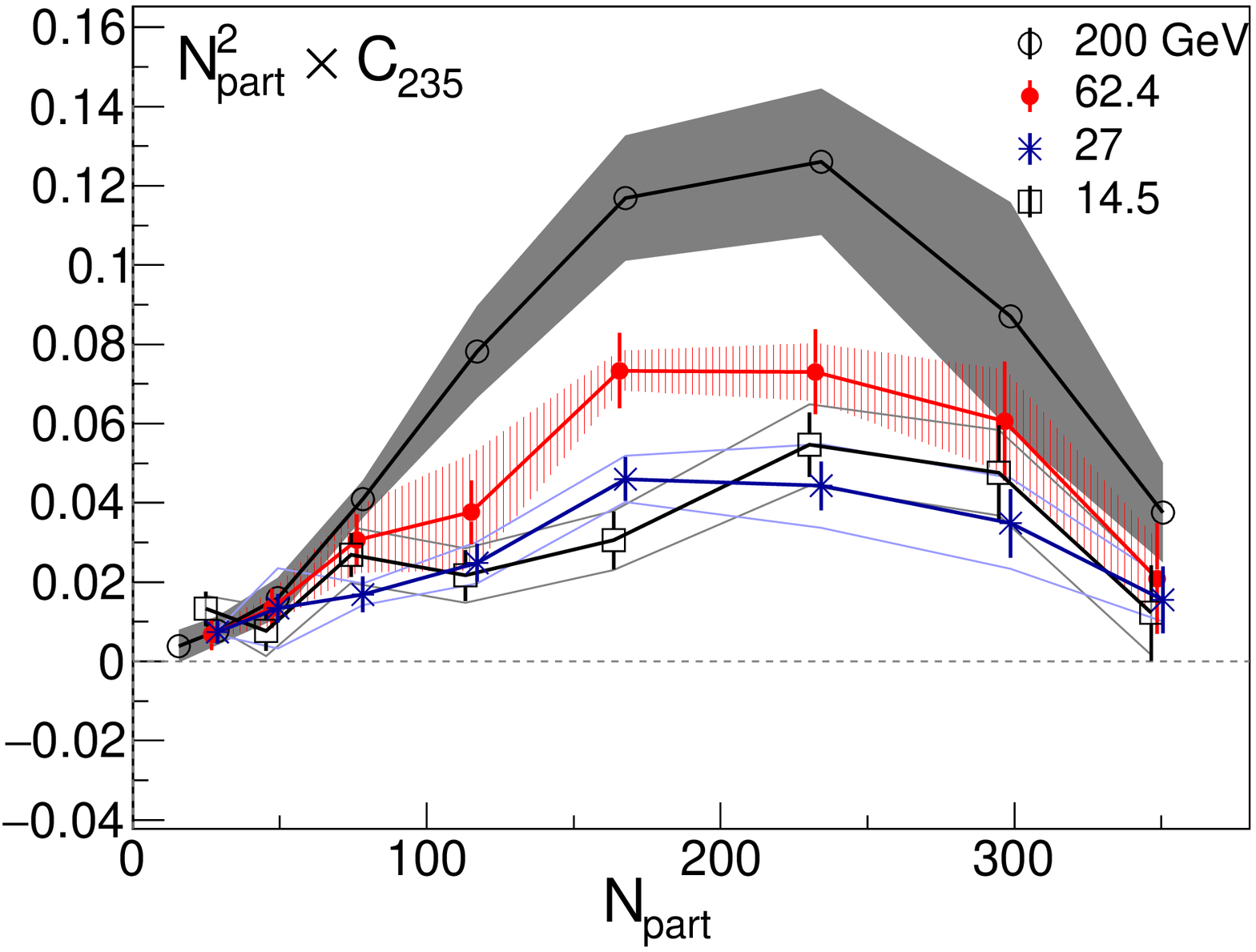}
\end{minipage}
\caption{ (color online) The centrality dependence of {\cbbd} (left)
  and {\cbce} (right) scaled by {\npartsq} for a selection of energies.  } \label{figb}
\end{figure*}

Figure~\ref{figb} shows the centrality dependence of {\npartsq}{\cbbd}
and {\npartsq}{\cbce} for a selection of collision energies. Both
{\cbbd} and {\cbce} remain positive for the centralities and energies
shown with no apparent changes in the centrality trends. We note
that although {\cbbd} drops significantly from 200 down to 19.6 GeV,
we observe little change with energy below 19.6 GeV. A similar lack of
energy dependence between 7.7 and 19.6 GeV was also observed in recent
measurements of $v_3^2\{2\}$~\cite{besv3}. This is notable since one
would naively expect either of these correlation measurements to
continuously increase as the density of the collision region increases.

\begin{figure*}[hbtp]
\centering\mbox{
\includegraphics[width=0.99\textwidth]{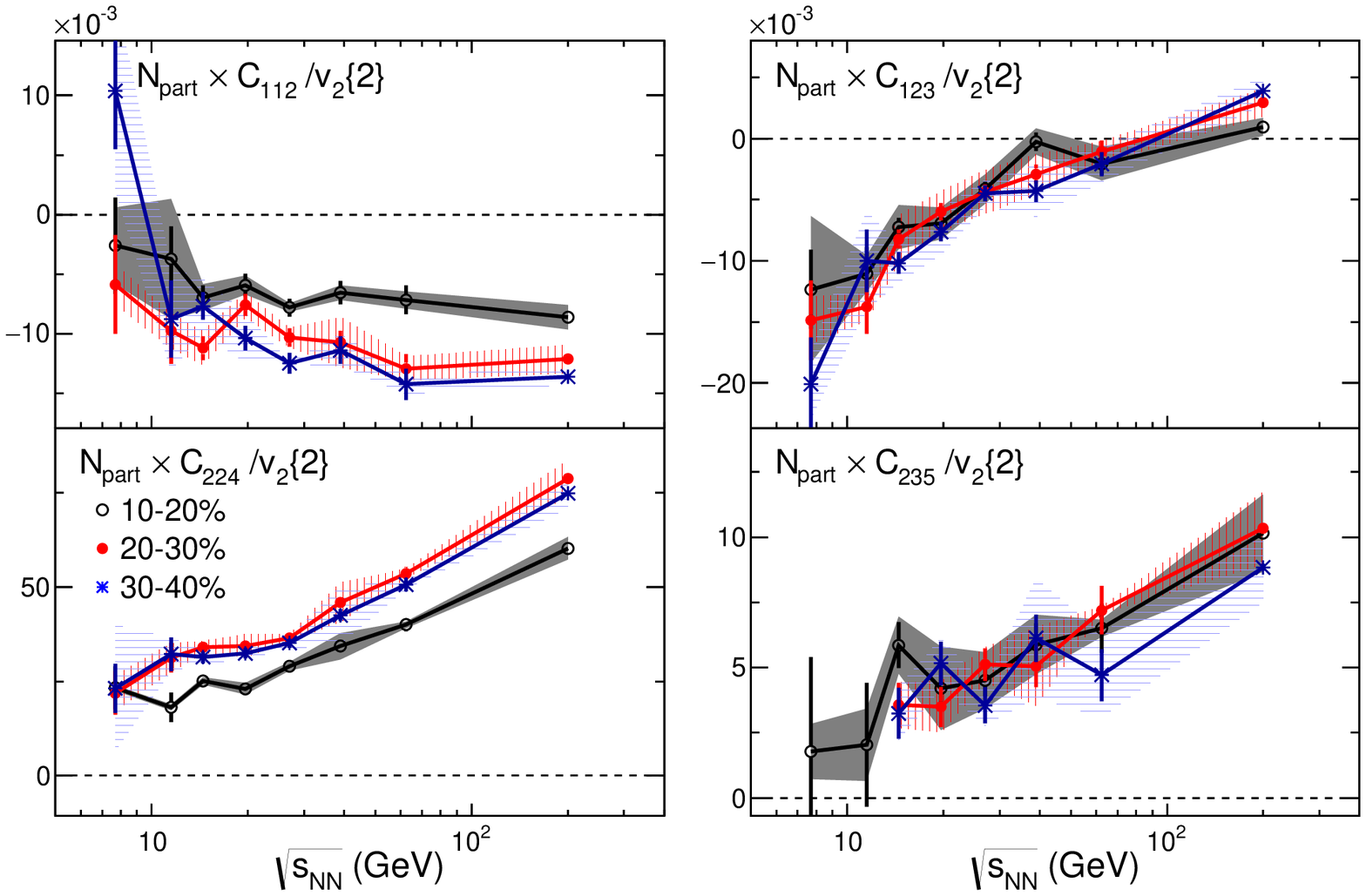}}
\caption{ (color online) The {\snn} dependence of
  {\npart}{\cijk}{/$v_2$} for $(m,n)=(1,1)$ (top left), $(1,2)$ (top
  right), $(2,2)$ (bottom left) and $(2,3)$ (bottom right) for three
  selected centrality intervals. In the bottom right panel, the lowest
  energy points for the 20-30\% and 30-40\% centrality intervals,
  having large uncertainties, are omitted for clarity. Statistical
  uncertainties are shown as vertical error bars while the systematic
  errors are shown as shaded regions or bands. } \label{edep}
\end{figure*}

To better view the energy trends, in Fig.~\ref{edep} we show
{\npart}{\cijk}$/v_2$ as a function of {\snn} for three centrality
intervals: 10-20\%, 20-30\%, and 30-40\%. The $v_2$ values are based
on a two-particle cumulant analysis as discussed in
Appendix~\ref{v2}. The scaling will be further discussed in the next
paragraph.  For all centrality intervals shown, {\caab}$/v_2$ is
negative at the highest energy but the magnitude of the correlation
decreases as the energy decreases and becomes consistent with zero,
although with large errors, at 7.7 GeV. This behavior was also
observed in the charge dependence of this correlator which has been
studied to search for the charge separation predicted to be a
consequence of the chiral magnetic effect~\cite{Adamczyk:2014mzf}. As
noted above, both {\cbbd} and {\cbce} are positive for all energies.
The energy dependence of {\cabc}$/v_2$ is unique in that it is
positive at 200 GeV but then drops below zero near 62.4 GeV and
continues to become more negative at lower energies. In the following paragraph, we discuss the implications that this trend has for how two-particle correlations with
respect to the reaction plane change with energy.

The correlations {\caab}, {\cabc}, {\cbbd}, and {\cbce} presented in Fig.~\ref{edep} have either $m=2$, $n=2$, or $m+n=2$. When $v_2$ is large, as it is for the 10-20\%, 20-30\% and 30-40\% centrality intervals, then $\langle\cos(1\phi_1+1\phi_2-2\phi_3)\rangle/v_2\approx \langle\cos(1\phi_1+1\phi_2-2\Psi_{\mathrm{RP}})\rangle$ and $\langle\cos(2\phi_1+m\phi_2-(m+2)\phi_3)\rangle/v_2\approx \langle\cos(2\Psi_{\mathrm{RP}}+m\phi_2-(m+2)\phi_3)\rangle$ where $\Psi_{\mathrm{RP}}$ is the reaction plane angle. Correlations including a second harmonic should then provide information about two-particle correlations with respect to the second harmonic reaction plane:
\begin{eqnarray}
  \langle\cos(1\phi_1+1\phi_3-2\phi_2)\rangle/v_{2} &\approx& \langle\cos(1\phi_1'+1\phi_2')\rangle, \nonumber \\
  \langle\cos(1\phi_1+2\phi_3-3\phi_2)\rangle/v_{2} &\approx& \langle\cos(1\phi_1'-3\phi_2')\rangle, \nonumber \\
  \langle\cos(2\phi_1+2\phi_3-4\phi_2)\rangle/v_{2} &\approx& \langle\cos(2\phi_1'-4\phi_2')\rangle, \nonumber \\
  \langle\cos(2\phi_3+3\phi_1-5\phi_2)\rangle/v_{2} &\approx& \langle\cos(3\phi_1'-5\phi_2')\rangle,
\label{eq3}
\end{eqnarray}
where $\phi'=\phi-\Psi_{\mathrm{RP}}$. Since we are integrating over all particles in these correlations, the subscript label for the particles is arbitrary so we have reassigned them so that particle 3 is always associated with the second harmonic. For illustration, Table~\ref{tab} shows values for {\cijk}$/v_2$ for specific values of $\phi_1'$ and $\phi_2'$. At 200 GeV, all measured correlations are positive except $\langle\cos(\phi_1'+\phi_2')\rangle$. This points to an enhanced probability for a pair of particles in one of two possible configurations: either $\phi_1'\approx \pi/3$ and $\phi_2'\approx 2\pi/3$ or $\phi_1' \approx -\pi/3$ and $\phi_2' \approx -2\pi/3$ (these correspond to the right-most column of Table~\ref{tab}). This result is surprising since it implies a preference for both of the correlated particles to either be in the upper hemisphere, or both in the lower hemisphere. We note however, that hydrodynamic models with fluctuating initial conditions correctly predict this trend~\cite{shortpaper} which could arise from increased density fluctuations at either the top or the bottom of the almond shaped overlap region. A high density fluctuation in the lower half of  the almond zone naturally leads to particles moving upward and away from that density fluctuation so that they both end up in the upper hemisphere. This response was described in Ref.~\cite{Teaney:2010vd} and was illustrated as ``Position B'' in Fig. IV of that reference. For energies below 200 GeV, {\cabc} changes sign so that $\langle\cos(\phi_1'+\phi_2')\rangle$ and $\langle\cos(1\phi_1'-3\phi_2')\rangle$ are both negative while $\langle\cos(2\phi_1'-4\phi_2')\rangle$ and $\langle\cos(3\phi_1'-5\phi_2')\rangle$ are both positive. This condition does not match any of the scenarios in the table but it could indicate an increased preference for particle pairs with $\phi_1' \approx 0$ and $\phi_2' \approx \pi$. A preference for back-to-back particle pairs aligned with the reaction plane would be consistent with an increased importance for momentum conservation at lower energies. Momentum conservation naturally leads to a tendency for particles to be emitted with back-to-back azimuth angles~\cite{Adamczyk:2013hsi}. As the beam energy is decreased, the multiplicity decreases and we should expect the effects of momentum conservation to become more prominent (in the case that only two particles are emitted, they must be back-to-back). The implications of this change in the configuration of two-particle correlations with respect to the reaction plane deserves further theoretical investigation.
\begin{table}[htb]
\caption{Values for {\cijk}/$v_2$ for specific cases of $\phi_1'$ and $\phi_2'$ where $\phi'=\phi-\Psi_{\mathrm{RP}}$ (see Eq.~\ref{eq3}). The first column ($\phi_1'=\phi_2'=0$) corresponds to a particle pair with $\Delta\phi=0$ emitted in the direction of the reaction plane (in-plane). The second column corresponds to back-to-back ($\Delta\phi=\pi$) particles emitted in-plane. The third and fourth columns correspond to pairs of particles emitted perpendicular to the reaction plane (out-of-plane) with either $\Delta\phi=0$ or $\Delta\phi=\pi$ respectively. The right-most column is a scenario consistent with the correlations observed in mid-central collisions at $\sqrt{s_{{\rm NN}}}=200$~GeV.}
\label{tab}
\begin{center}
\bgroup
\def\arraystretch{1.4}
\begin{tabular}{ l c c c c c}
\hline
\hline
& \multicolumn{5}{c}{($\phi_1',~\phi_2'$) [rad] } \\
\cline{2-6}
& ~(0,~0)~ & ~(0,~$\pi$)~ & $\pm(\frac{\pi}{2},~\frac{\pi}{2}$)~ & ($\frac{\pi}{2},-\frac{\pi}{2}$)~ & $\pm(\frac{\pi}{3},~\frac{2\pi}{3})$~ \\ 
\hline
 {\caab}/$v_2$ & +1 & -1 & -1 & +1 & -1 \\  
 {\cabc}/$v_2$ & +1 & -1 & -1 & +1 & +$\frac{1}{2}$ \\  
 {\cbbd}/$v_2$ & +1 & +1 & -1 & -1 & +1 \\  
 {\cbce}/$v_2$ & +1 & -1 & -1 & +1 & +$\frac{1}{2}$\\
\hline
\hline
\end{tabular}
\egroup
\end{center}
\end{table}

The discussion in the above paragraph illustrates how measurements of {\cijk} reveal information about two-particle correlations with respect to the reaction plane and we pointed out two specific conclusions based on the $p_T$- and $\Delta\eta$-integrated measurements. The value of {\cabc} changes sign as a function of centrality, $\Delta\eta$ and $p_T$ suggesting that further specific configurations may arise when triggering on a particular $p_T$ or investigating particles separated by an $\eta$-gap. We have not examined the charge dependence of {\cijk} but future work placing a like-sign or unlike-sign requirement on $\phi_1'$ and $\phi_2'$ may be useful for interpreting charge separation measurements and determining whether they should be taken as evidence for the chiral magnetic effect.

\section{Conclusions}
\label{co}

We presented measurements of the energy, centrality, $p_T$, and $\Delta\eta$ dependence of three-particle azimuthal correlations {\cijk} for a variety of combinations of $m$ and $n$. We find a strong dependence of {\caab} on $|\eta_1-\eta_2|$ and a strong dependence of {\cabc} on $|\eta_1-\eta_3|$. Meanwhile, {\cbbd} and {\cbce} exhibit a smaller but still appreciable dependence on $|\eta_{1}-\eta_3|$. This may indicate either the presence of short-range non-flow correlations or a rapidity dependence to the initial energy density signaling a breaking of longitudinal invariance. Simple pictures of non-flow however, appear to be inconsistent with the overall trends observed in the data. The integrated correlations with $m=1$ are generally negative or consistent with zero except for {\cabc} which, at 200 GeV, is positive for mid-central collisions while it is negative for all centralities at all of the lower energies. Nonzero values for {\cabc} imply correlations between the second and third harmonic event plane that are predicted from models of the initial overlap geometry. 
The $p_T$ dependence of the correlations exhibits trends suggesting significant differences between the correlations in peripheral collisions and more central collisions as well as differences for $p_T>5$ GeV/$c$ and $p_T<5$ GeV/$c$. The quantity {\cabc} as a function of $p_{T,1}$ changes sign as many as three times. While {\caab} is negative for higher energies, it becomes positive or consistent with zero at 7.7 GeV. By examining the energy dependence of {\caab}, {\cabc}, {\cbbd}, and {\cbce} divided by $v_2$ we are able to infer that in mid-central collisions at 200 GeV, there is a preference for particle pairs to be emitted with angles relative to the reaction plane of either $\phi_1\approx\pi/3$ and $\phi_2\approx2\pi/3$ or $\phi_1\approx-\pi/3$ and $\phi_2\approx-2\pi/3$. At 62.4 GeV and below, this appears to change due to a possible preference for back-to-back pairs ($\phi_1\approx0$ and $\phi_2\approx\pi$) aligned with the reaction plane. These data will be useful for constraining hydrodynamic models~\cite{shortpaper}. In order to facilitate such future data-model comparisons we also include the measurements of $v_{n}^2\{2\}, n=1,2,4,5$, over a wide range of energy, in the appendix of this paper. Measurements of the charge dependence of the correlations presented here, by revealing information about the preferred directions of correlated particles with respect to the reaction plane, should provide valuable insights into whether or not the charge separation observed in heavy-ion collisions is related to the chiral magnetic effect.

\section{Summary}
\label{su}
The very first measurement of charge inclusive three-particle azimuthal correlations from the RHIC beam energy scan program, presented in this paper, can provide several new insights into the initial state and transport in heavy ion collisions. These observables go beyond conventional flow harmonics and provide the most efficient way of studying the correlation between harmonic amplitudes and their phases over a wide range of multiplicities. These observables are well defined and of general interests even when the azimuthal correlations are not dominated by hydrodynamic flow. The major finding of this analysis is the strong relative pseudorapidity ($\Delta\eta$) dependence between the particles associated with different harmonics, observed up to about two units ($\Delta\eta\sim$~2) of separation. Non-flow based expectations such as fragmentation ($\Delta\eta\sim$~1) or momentum conservation (flat in $\Delta\eta$) can not provide a simple explanation to such observations. If the observed correlations are dominated by flow, the current results strongly hint at a breaking of longitudinal invariance of the initial state geometry at RHIC. The comprehensive study of momentum and centrality dependence of three-particle correlations over a wide range of energy (7.7-200 GeV), presented here, will help reduce the large uncertainties in the transport parameters involved in hydrodynamic modeling of heavy ion collisions over a wide range of temperature and net-baryon densities. In addition, the charge inclusive three-particle correlations will provide baselines for the measurements of the chiral magnetic effect.  


\section*{Acknowledgments}
We thank the RHIC Operations Group and RCF at BNL, the NERSC Center at LBNL, and the Open Science Grid consortium for providing resources and support. This work was supported in part by the Office of Nuclear Physics within the U.S. DOE Office of Science, the U.S. National Science Foundation, the Ministry of Education and Science of the Russian Federation, National Natural Science Foundation of China, Chinese Academy of Science, the Ministry of Science and Technology of China and the Chinese Ministry of Education, the National Research Foundation of Korea, GA and MSMT of the Czech Republic, Department of Atomic Energy and Department of Science and Technology of the Government of India; the National Science Centre of Poland, National Research Foundation, the Ministry of Science, Education and Sports of the Republic of Croatia, RosAtom of Russia and German Bundesministerium fur Bildung, Wissenschaft, Forschung and Technologie (BMBF) and the Helmholtz Association.

\appendix
\section{Two-particle Cumulants $v_{n}^{2}\{2\}$}
\label{v2}

In this appendix we present the measurements of $v_{n}^2\{2\}$ for
n=1, 2, 4 and 5. The second harmonic $v_{2}^2\{2\}$ was used to
scale {\cijk} in Fig.~\ref{edep}. Under the assumption that
\begin{eqnarray}
\label{eq_eplane}
\langle \cos(m\phi_1 &+& n \phi_2 - (m+n) \phi_3)\rangle \approx \\
&&\!\!\!\!\!\!\!\!\!\!\!\! \langle v_m v_n v_{m+n}\cos(m \Psi_m + n \Psi_n - (m+n) \Psi_{m+n}) \rangle \nonumber
\end{eqnarray}
where $\Psi_m$ is the participant plane angle for harmonic $m$, one
can convert the {\cijk} correlations into reaction plane correlations in the low-resolution limit
by dividing by $\sqrt{v_m^2\{2\}v_n^2\{2\}v_{m+n}^2\{2\}}$. The relationship of the {\cijk} to $v_m$ and $\Psi_m$ assumes that non-flow correlations are minimal. Similar assumptions must also be made when using event-plane angles in the analysis. The
analysis of $v_{n}^2\{2\}$ was performed in a similar manner to that
of $v_{3}^2\{2\}$ presented in Ref.~\cite{besv3}. The $\Delta\eta$
dependence of $\langle\cos2(\phi_1-\phi_2)\rangle$ is analyzed for
$p_T>0.2$ GeV/$c$ and $|\eta|<1$. Short-range correlations are
parameterized with a narrow Gaussian peak centered at $\Delta\eta=0$
and the remaining longer-range correlations are integrated (weighting
by the number of pairs at each $\Delta\eta$) to obtain the
$\Delta\eta$-integrated $v_n^2\{2\}$ results. The quantity labeled
$v_2$ in Fig.~\ref{edep} is $\sqrt{v_2^2\{2\}}$.

\begin{figure*}[!hbtp]
\centering
\begin{minipage}{0.495\textwidth}
\centering
\includegraphics[width=1.0\textwidth]{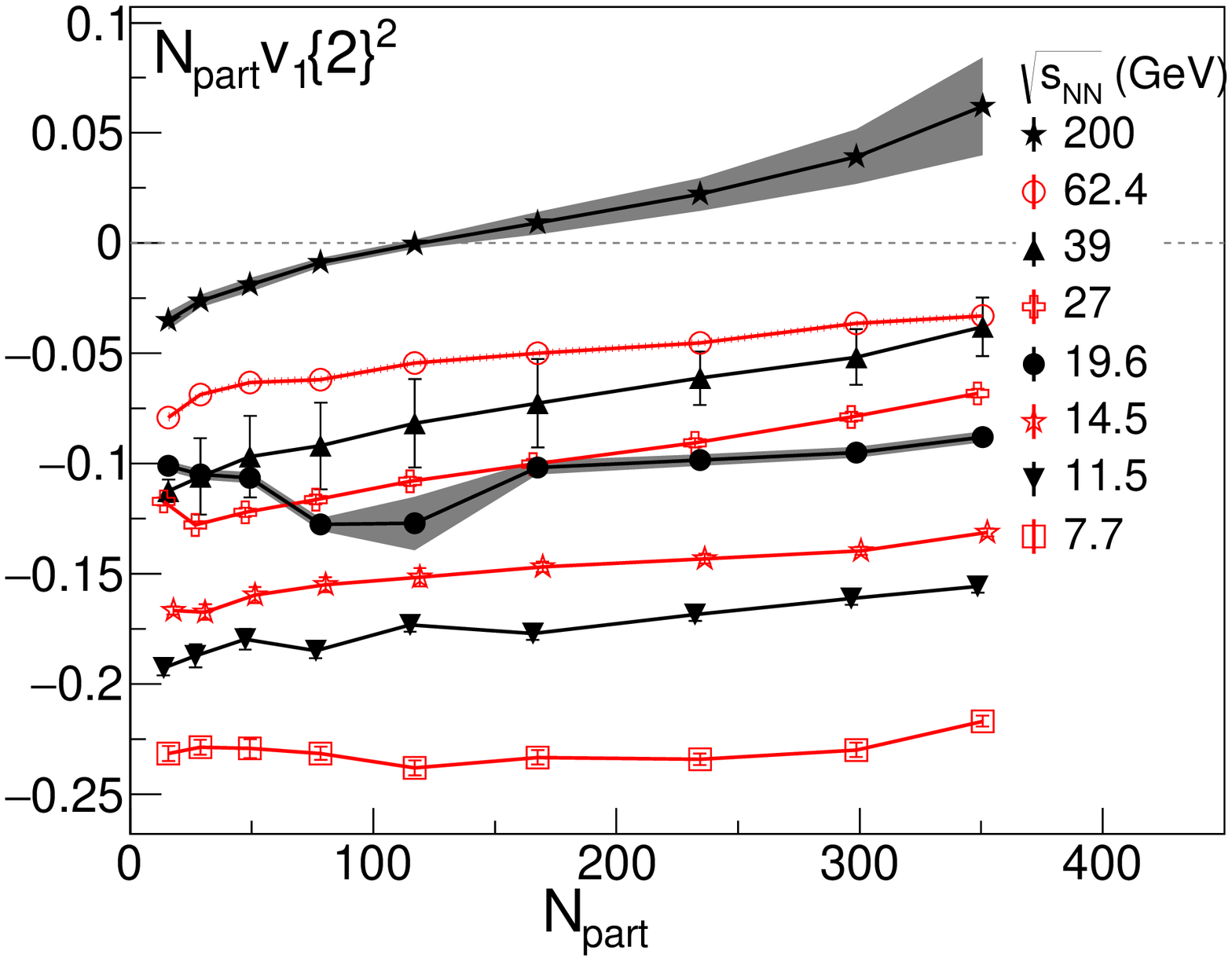}
\end{minipage}
\begin{minipage}{0.495\textwidth}
\centering
\includegraphics[width=1.0\textwidth]{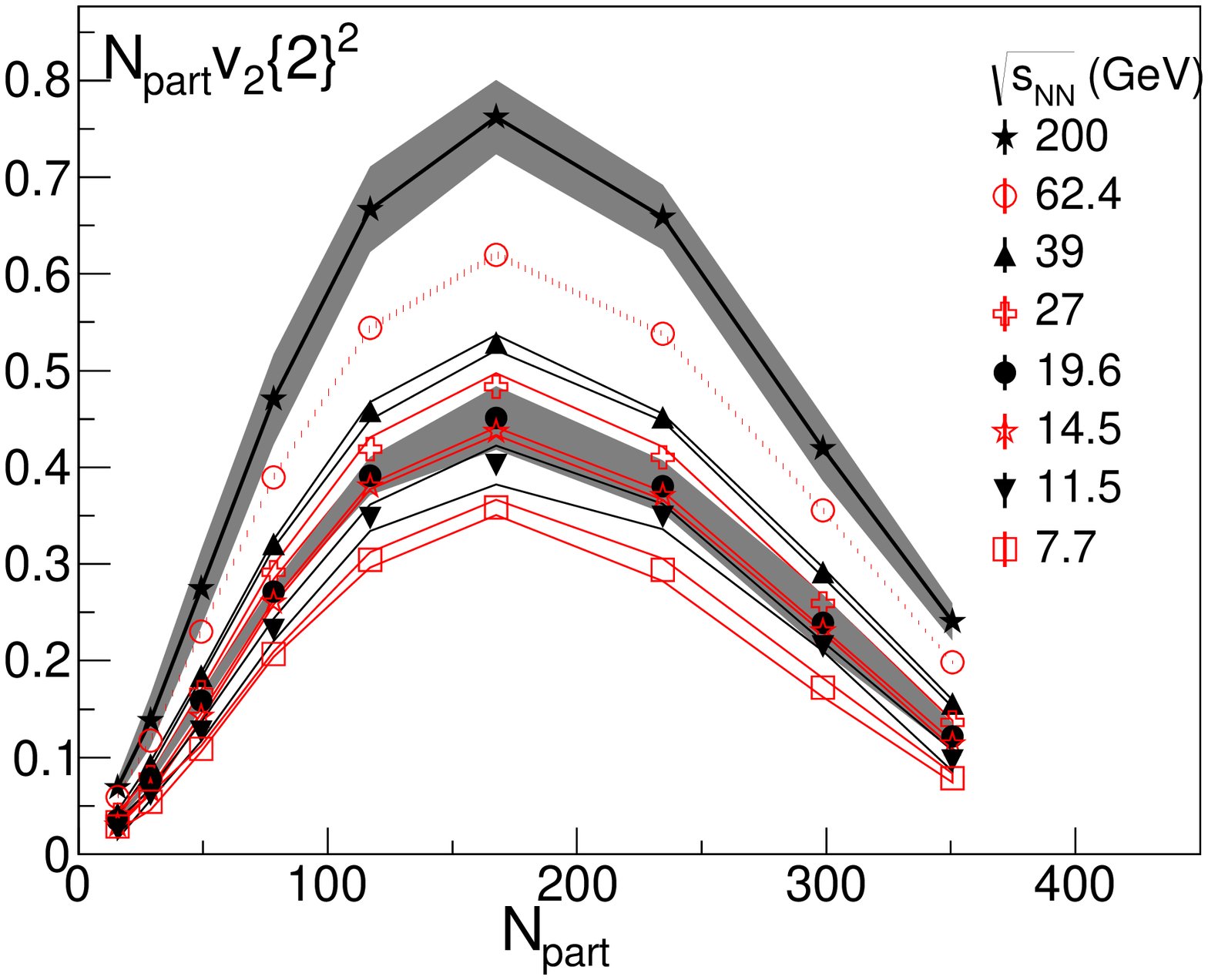}
\end{minipage}
\caption{  The {\snn} dependence and centrality
  dependence of {\npart}{$v_1^2\{2\}$} (left) and {\npart}{$v_2^2\{2\}$} (right) after short-range correlations,
  predominantly from quantum and Coulomb effects, have been
  subtracted. For more details see Ref.~\cite{besv3}. The centrality
  intervals correspond to 0-5\%, 5-10\%, 10-20\%, 20-30\%, 30-40\%,
  40-50\%, 50-60\%, 60-70\% and 70-80\%. The {\npart} values used for
  the corresponding centralities are 350.6, 298.6, 234.3, 167.6,
  117.1, 78.3, 49.3, 28.2 and 15.7 independent of energy.
} \label{v12}
\end{figure*}

Figure~\ref{v12} shows the results for $v_1^2\{2\}$ (left) and
$v_2^2\{2\}$ (right) as a function of centrality for 200, 62.4, 39,
27, 19.6, 14.5, 11.5, and 7.7 GeV Au+Au collisions. The data are
scaled by $N_{\mathrm{part}}$ and plotted verses $N_{\mathrm{part}}$ for
convenience. At 200 GeV, $v_1^2\{2\}$ is positive for central
collisions but becomes negative for {\npart}$<150$. The negative values
are expected from momentum conservation and present a conceptual
challenge for dividing {\cijk} by $\sqrt{v_1^2\{2\}}$. The values of
$v_1^2\{2\}$ become more negative at lower energies. This is
consistent again with momentum conservation effects which are expected
to become stronger as multiplicity decreases. In the limit of a
collision that produces only two particles, momentum conservation
would require that $v_1^2\{2\}=-1$. The $v_1^2\{2\}$ results follow a
monotonic energy trend except for peripheral collisions at 19.6 GeV
which appear to be elevated with respect to the trends.

The right panel of Fig.~\ref{v12} shows the results for
{\npart}$v_2^2\{2\}$ which remain positive for all energies and
collision centralities. While it is unusual to scale $v_2^2\{2\}$ by
{\npart}, we keep this format for consistency. The scaled results
exhibit a strong peak for mid-central collisions due to the elliptic
geometry of those collisions.

\begin{figure*}[!hbtp]
\centering
\begin{minipage}{0.495\textwidth}
\centering
\includegraphics[width=1.0\textwidth]{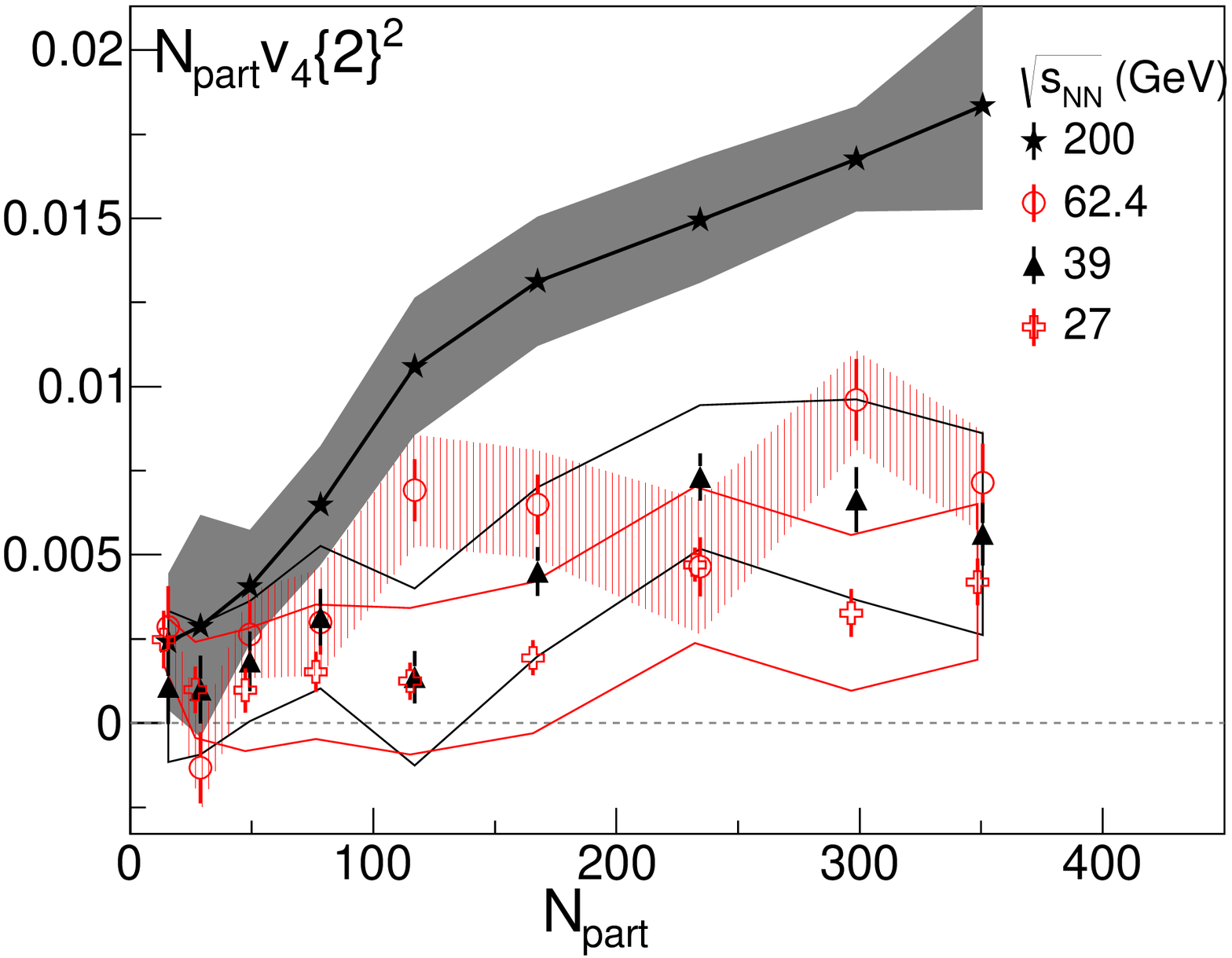}
\end{minipage}
\begin{minipage}{0.495\textwidth}
\centering
\includegraphics[width=1.0\textwidth]{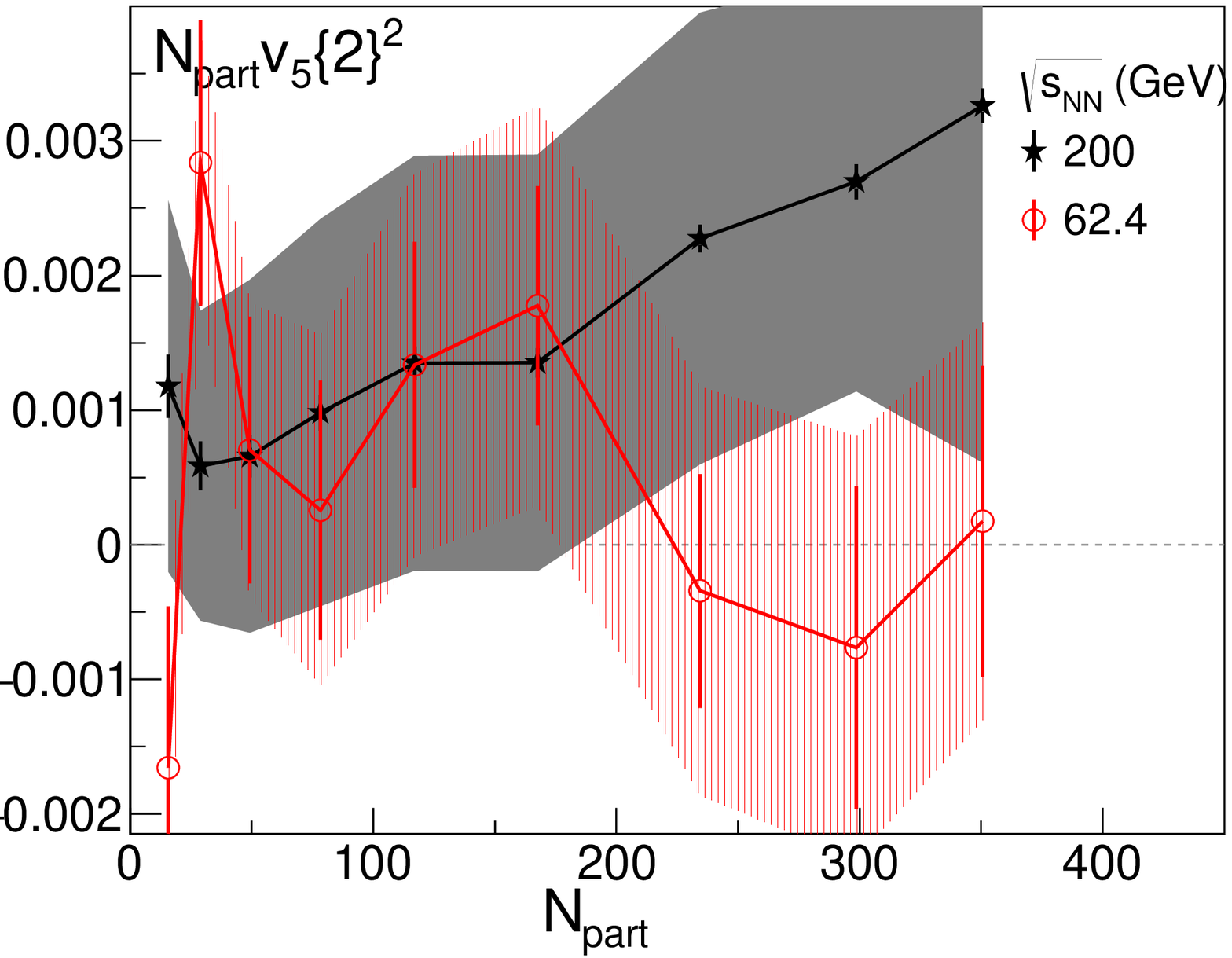}
\end{minipage}
\caption{  The {\snn} dependence and centrality
  dependence of {\npart}{$v_4^2\{2\}$} (left) and {\npart}{$v_5^2\{2\}$} (right) after short-range correlations,
  predominantly from Quantum and Coulomb effects, have been
  subtracted. For more details see Ref.~\cite{besv3}.
} \label{v45}
\end{figure*}

Figure~\ref{v45} shows the data for {\npart}$v_4^2\{2\}$ (left) and
{\npart}$v_5^2\{2\}$ (right) for a more limited energy range. Results for
{\npart}$v_3^2\{2\}$ are available in Ref.~\cite{besv3}. At the lower
energies the relative uncertainties on these data become too large to
be of use. This presents another challenge to recasting {\cijk} in
terms of reaction plane correlations because scaling by $\sqrt{v_4^2\{2\}}$ or
$\sqrt{v_5^2\{2\}}$ leads to a large uncertainty on the resulting
ratios.

\end{document}